\documentclass{article}
\usepackage[utf8]{inputenc}
\usepackage{enumerate}

\usepackage{appendix}
\usepackage{mathrsfs}
\usepackage{amsmath}
\usepackage{amssymb}
\usepackage{float}
\usepackage{graphicx}
\usepackage{multirow}
\usepackage{xcolor}
\usepackage{bm}
\usepackage[normalem]{ulem}
\usepackage[autostyle]{csquotes}

\usepackage{booktabs}
\usepackage{color}
\usepackage{jcappub}
\newcommand{\be}{\begin{equation}}
\newcommand{\ee}{\end{equation}}
\newcommand{\bea}{\begin{eqnarray}}
\newcommand{\eea}{\end{eqnarray}}

\DeclareMathOperator{\sgn}{sgn}

\renewcommand{\vec}[1]{ {\mathbf #1} } 

\def\ltsima{$\; \buildrel < \over \sim \;$}
\def\simlt{\lower.5ex\hbox{\ltsima}}
\def\gtsima{$\; \buildrel > \over \sim \;$}
\def\simgt{\lower.5ex\hbox{\gtsima}}

\begin{document}
\title{Cosmological simulations of scale-dependent primordial non-Gaussianity}
\author[1,2,3]{Marco Baldi,}
\author[4]{Emanuele Fondi,}
\author[5,6]{Dionysios Karagiannis,}
\author[1,2,3]{Lauro Moscardini,}
\author[7,8,4]{Andrea Ravenni,}
\author[9,10]{William R. Coulton,}
\author[11]{Gabriel Jung,}
\author[7,8]{Michele Liguori,}
\author[7,8]{Marco Marinucci,}
\author[4,12]{Licia Verde,}
\author[13]{Francisco Villaescusa-Navarro,}
\author[14,13]{Benjamin D.~Wandelt}

\affiliation[1]{Dipartimento di Fisica e Astronomia, Alma Mater Studiorum Universit\`a di Bologna, via Piero Gobetti, 93/2, I-40129 Bologna, Italy;}
\affiliation[2]{Osservatorio di Astrofisica e Scienza dello Spazio, via Piero Gobetti 93/3 1, I-40129 Bologna, Italy;}
\affiliation[3]{INFN - Sezione di Bologna, viale Berti Pichat 6/2, I-40127 Bologna, Italy;}
\affiliation[4]{Instituto de Ciencias del Cosmos, University of Barcelona, ICCUB, Mart\' i i Franqu\` es, 1, E08028 Barcelona, Spain}
\affiliation[5]{Department of Physics \& Astronomy, Queen Mary University of London, London E1 4NS, United Kingdom}
\affiliation[6]{Department of Physics \& Astronomy, University of the Western Cape, Cape Town 7535, South Africa}
\affiliation[7]{Dipartimento di Fisica e Astronomia “G. Galilei”, Università degli Studi di Padova, via Marzolo 8, I-35131, Padova, Italy}
\affiliation[8]{INFN, Sezione di Padova, via Marzolo 8, I-35131, Padova, Italy}
\affiliation[9]{Kavli Institute for Cosmology Cambridge, Madingley Road, Cambridge CB3 0HA, UK}
\affiliation[10]{DAMTP, Centre for Mathematical Sciences, University of Cambridge, Wilberforce Road, Cambridge CB3 OWA, UK}
\affiliation[11]{Université Paris-Saclay, CNRS, Institut d’Astrophysique Spatiale, 91405 Orsay, France}
\affiliation[12]{ICREA, Pg. Lluis Companys 23, Barcelona, 08010, Spain}
\affiliation[13]{Center for Computational Astrophysics, 160 5th Avenue, New York, NY, 10010, USA}
\affiliation[14]{Sorbonne Universit\'{e}, CNRS, UMR 7095, Institut d'Astrophysique de Paris, 98 bis bd Arago, 75014 Paris, France}

\emailAdd{marco.baldi5@unibo.it}

\abstract{We present the results of a set of cosmological N-body simulations with standard $\Lambda $CDM cosmology but characterized by a scale-dependent primordial non-Gaussianity of the {\em local} type featuring a power-law dependence of the $f_{\rm NL}^{\rm loc}(k)$ at large scales followed by a saturation to a constant value at smaller scales where non-linear growth leads to the formation of collapsed cosmic structures. Such models are built to ensure consistency with current Cosmic Microwave Background bounds on primordial non-Gaussianity yet allowing for large effects of the non-Gaussian statistics on the properties of non-linear structure formation. We show the impact of such scale-dependent non-Gaussian scenarios on a wide range of properties of the resulting cosmic structures, such as the non-linear matter power spectrum, the halo and sub-halo mass functions, the concentration-mass relation, the halo and void density profiles, and we highlight for the first time that some of these models might mimic the effects of Warm Dark Matter for several of such observables.}

\maketitle
\flushbottom
%\begin{keywords}
%dark energy -- dark matter --  cosmology: theory -- galaxies: formation
%\end{keywords}

%*****************************************************************************

\section{Introduction}
\label{i}

The unprecedented precision of on-going  or forthcoming  observations in cosmology holds the potential to  measure cosmological observables and constrain cosmological parameters with percent-level (statistical) uncertanties.  Hence the  modeling of the different physical processes that drive the evolution of the universe and its observable properties must attain a comparable level of accuracy.

In particular, state-of the-art or imminent 
cosmological surveys  -- such as e.g. {\em Euclid} \citep{Laureijs_etal_2011,Euclid:2021icp,2024arXiv240513491E}; the Dark Energy Spectroscopic Instrument \citep[DESI, ][]{DESI:2016fyo}, the \textit{Vera C. Rubin} Observatory Legacy Survey of Space and Time \citep[LSST, ][]{LSST:2008ijt}, the Spectro-Photometer for the History of the Universe, Epoch of Reionization, and the Ices Explorer \citep[SPHEREx, ][]{SPHEREx:2014bgr}, and the \textit{Nancy Grace Roman} Space Telescope \citep{Spergel:2015sza} -- 
will focus on the  
 evolved universe at low redshifts.
 Therefore, possible signatures of primordial physical processes, such as e.g. the detailed footprints of different inflationary models \citep[][]{Planck:2019kim,Euclid_TWG_2}, will have to be extracted from the highly-evolved distribution of large-scale structures.  As a consequence of the 
 non-linear evolution of the initial density field  
 through gravitational instability, 
 these structures may retain only a feeble reminiscence of the statistical properties of their primordial seeds. Furthermore, for surveys that exploit visible galaxies as the tracers of the underlying matter distribution, the complex relations between different galaxy observable properties and their clustering bias, large-scale environment, and formation history must be properly modeled and taken into account in order to be able to extract any meaningful cosmological information from the data \citep[][]{Euclid:2024few}. 

In this respect, the use of numerical N-body simulations \citep[see e.g. ][for a general review]{Kuhlen_Vogelsberger_Angulo_2012,Angulo:2021kes} has become an essential ingredient for the preparation and the exploitation of most upcoming surveys. 
Simulated data of key observables quantities are essential  
to test  
performance, 
accuracy, and 
reliability of the different statistical analysis tools involved in  extracting information about standard cosmological parameters and in interpreting possible hints of new physics beyond the standard model. Cosmological simulations are also of primary importance in modeling possible extensions of the standard cosmological scenario, as e.g. Dark Energy \citep[][]{Baldi_2012b}, Modified Gravity \citep[][]{Winther_etal_2015}, massive neutrinos \citep[][]{Euclid:2022qde}, non-thermal Dark Matter particle candidates \citep[][]{Nori:2018hud}, and to accurately predict their observational footprints deep into the non-linear regime of structure formation also (possibly) including sophisticated sub-grid implementations of baryonic physics and astrophysical processes \citep[][]{Vogelsberger:2019ynw,Arnold:2019vpg}.\\

This is particularly relevant for models of primordial non-Gaussianity, which will be the focus of the present work. From a theoretical point of view, {some level of primordial non-Gaussianity is a general prediction of nearly all inflationary models} \citep[][]{Liguori:2003mb,Bartolo_etal_2004}. Observationally, the magnitude  
of such deviation from Gaussianity  
has been tightly constrained by the last two decades of Cosmic Microwave background (CMB) results, such as the {\em WMAP} \citep[][]{WMAP:2008lyn,wmap7,wmap9} and the {\em Planck} \citep[][]{Planck_2013_NG,Planck_2015_NG,Planck:2019kim} collaborations, {with upper values that allowed} ruling out a wide range of inflationary scenarios, and consequently making a possible observational detection of primordial non-Gaussianity {from low-redshift observations} extremely challenging. Nonetheless, such observational bounds have been derived under the assumption of a scale-independent non-Gaussianity, and can be evaded by considering models where such assumption does not hold.

In particular, in the present work we will consider cosmological scenarios characterized by a primordial density field with non-Gaussian statistics described by local-type non-Gaussianity with a scale-dependent amplitude $f_{\rm NL}(k)$ \citep[][]{Sefusatti_etal_2009, Oppizzi_etal_2018} resulting in a large non-Gaussianity at small scales  while still matching current large-scale observational bounds. 
Recently, \cite{Stahl:2024stz} used a functional form  for $f_{\rm NL}(k)$ proportional to  $1+\tanh$ -- yielding a non-vanishing non-Gaussianity at all scales --  and  focussed on the dark matter power spectrum. 
Here we use a pure $\tanh$ functional form  and 
test extensively  the impact that such scale-dependent non-Gaussianity has 
on a wider range of statistics of the evolved matter density field at low redshifts, including statistical properties of biased tracers, both for positive and negative non-Gaussianity.  We also
highlight {for the first time} an intriguing similarity with the effects of Warm Dark Matter characterising models with a negative (and large, in absolute value) non-Gaussianity parameter at small scales. This unexpected outcome represents a further example of observational degeneracies between independent extensions of the standard cosmological scenario, following the previously investigated {\em cosmic degeneracies} between e.g. Modified Gravity and massive neutrinos \citep[see ][]{Baldi_etal_2014,Giocoli_Baldi_Moscardini_2018}, Modified Gravity and Warm Dark Matter \citep[][]{Baldi_Villaescusa-Navarro_2018}, and primordial non-Gaussianity and Dark Energy \citep[][]{Hashim_etal_2014} models.\\

The paper is organized as follows. In Section~\ref{sec:mods} we introduce the cosmological models under investigation. In Section~\ref{sec:simulations} we describe the numerical approach adopted to run collisionless cosmological N-body simulations for such models, and the data products extracted from the simulations. In Section~\ref{sec:results} we present the results of our analysis focusing on different observables extracted from the simulations, and in Section~\ref{sec:conclusions} we finally draw our conclusions.

\section{Cosmological Models}
\label{sec:mods}

\begin{table}
\begin{center}
\begin{tabular}{cc}
\hline
\hline
$\Omega_{\rm M}$ & 0.3175\\
$\Omega_{\Lambda }$ & 0.6825\\
$\Omega_{\rm b}$ & 0.049\\
$H_{0}$ & $67.11$ km~$s^{-1}$~ ${\rm Mpc}^{-1}$\\
$n_{s}$ & $0.96$\\
$A_{s}$ & $2.215\times 10^{-9}$ \\
\hline
\hline
\end{tabular}
\end{center}
\caption{The cosmological parameters consistent with the 2015 {\em Planck} data release \citep[][]{Planck_2015_XIII} 
adopted in the simulations presented in this work.}
\label{tab:params}
\end{table} 

We consider a set of cosmological models within the standard $\Lambda $CDM scenario, sharing the same set of cosmological parameters -- consistent with the 5-year constraints from the {\em Planck} satellite mission \citep[][]{Planck_2015_XIII} and summarised in Table~\ref{tab:params} -- but having different statistics of the primordial density perturbations. More specifically, we 
consider  models characterized by a non-Gaussian distribution of the primordial gravitational potential seeded by inflation.  Because deviations from Gaussianity are cosntrained to be small,  virtually all models expand the gravitational potential $\Phi$ into a Gaussian component $\Phi^G$ and a non-Gaussian correction $\Phi^{NG}$ whose amplitude is governed by a non-Gaussian parameter usually referred to as $f_{\rm NL}$.

The most widely studied primordial non-Gaussianity is that of the {\em local} type: the gravitational potential has the form
\begin{equation}
\Phi  
(\vec{x})= \Phi ^{\rm G}(\vec{x}) + f_{\rm NL}^{\rm loc}\left( \Phi ^{\rm G}(\vec{x})^{2} - \langle \Phi ^{\rm G}(\vec{x})^{2}\rangle \right) \,,
\label{eq:localNG}
\end{equation}
where $\Phi ^{\rm G}(\vec{x})$ is the real-space Gaussian potential. {As already mentioned above,} observational constraints on the non-Gaussianity parameter $f_{\rm NL}^{\rm loc}$ have improved over the past decades from the $f_{\rm NL}^{\rm loc} = 37.2 \pm 19.9$ of the {\em WMAP} final release \citep[][]{wmap9} to the much tighter bound $f_{\rm NL}^{\rm loc} = -0.9 \pm 5.1$ of the latest {\em Planck} release \citep[][]{Planck:2019kim}. Although no theoretical floor exists for the $f_{\rm NL}$ parameter, so that its actual value could be arbitrarily small and yet not identically vanishing, the simplest slow-roll single-field inflationary models predict $f_{\rm NL}\approx 10^{-3}$ \citep[][]{Cabass:2016cgp}, which appears to be significantly beyond the current observational bounds. 

\begin{figure}
\center
\includegraphics[width=0.7\columnwidth]{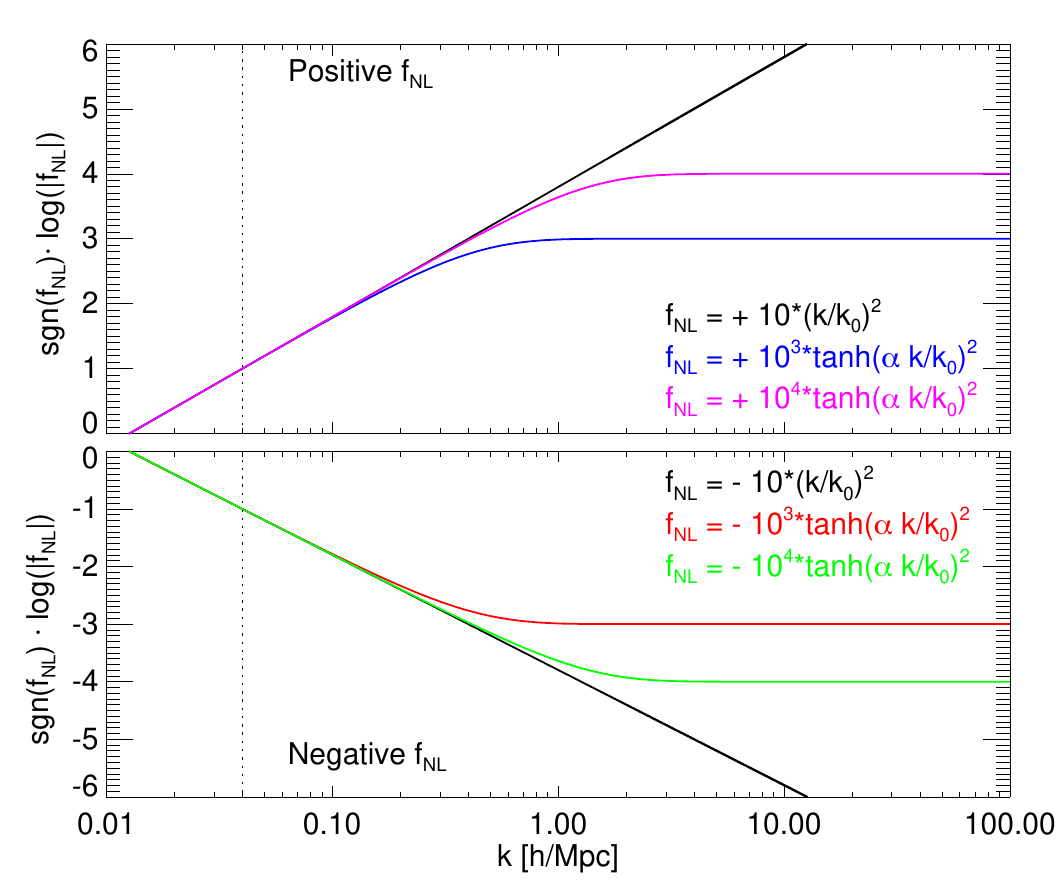}
\caption{The evolution of the $f_{\rm NL}^{\rm loc}(k)$ function for the models under investigation as a function of scale. To distinguish positive ({\em top panel}) and negative ({\em bottom panel}) non-Gaussianity we plot the quantity $\sgn (f_{\rm NL}^{\rm loc})\cdot \log (|f_{\rm NL}^{\rm loc}|)$. The vertical dotted line represents the pivot scale adopted by the {\em Planck} collaboration.}
\label{fig:models}
\end{figure}

Interestingly, however, the above-mentioned observational bounds have been derived by constraining $f_{\rm NL}$ at quite large scales (typically well within the linear regime down to $z=0$) and assuming it to be scale-independent. 
However, more complex phenomena or the presence of multiple fields during inflation may lead to larger values \citep[][]{Byrnes:2010em} or to a scale-dependence \citep[see e.g.][and references therein]{Sefusatti_etal_2009,Byrnes:2010ft,Wands:2010af,Stahl:2024stz} of the $f_{\rm NL}$ parameter. This has raised a significant interest in scale-dependent (also known as {\em running}) non-Gaussianity scenarios and their possible resulting phenomenology. { Scale-dependent primordial  non-Gaussianity is almost exclusively studied in Fourier space where  $f_{\rm NL}^{\rm loc}$  in the Fourier counterpart of  Eq. \ref{eq:localNG}   acquires a dependence on the wavenumber $k$.} The simplest and most widely considered form of {\em running} non-Gaussianity is characterized by an $f_{\rm NL}(k)$ 
of  the form \citep[][]{Sefusatti_etal_2009}:
\begin{equation}
f_{\rm NL}(k) = f_{\rm NL, 0} \left( \frac{k}{k_{0}} \right)^{n_{\rm NG}} 
\label{eq:runningfNL}
\end{equation}
where $n_{\rm NG}$ is the {\em running} index. If $k_{0}$ is taken to be the {\em pivot} scale at which CMB constraints are derived (i.e. $k_{0}=0.002$ Mpc$^{-1}$ for {\em WMAP} and $k_{0}=0.05$ Mpc$^{-1}$ for {\em Planck}), and $f_{\rm NL, 0}$ to be consistent with CMB bounds, a running index $n_{\rm NG}>0$ would result in stronger non-Gaussianity at smaller scales.
Figure~\ref{fig:models} shows the dependence on scale of $f_{\rm NL}^{\rm loc}$ as parameterised in Eq.~\ref{eq:runningfNL} (solid black curves) for a {\em running} index $n_{\rm NG}=2$ for both positive ({\em top panel}) and negative ({\em bottom panel}) non-Gaussianity. In the plots, the vertical dotted lines indicate the pivot scale $k_{0}=0.05$ Mpc$^{-1}$ assumed by \cite{Planck:2019kim} such that the choice $|f_{\rm NL}^{\rm loc}(k_{0})| \lesssim 10$ ensures $\approx 2\sigma$ consistency with current observational constraints, still allowing $|f_{\rm NL}^{\rm loc}|$ to increase way above such value at scales $k>k_{0}$. Models of such type have been extensively investigated in the literature \citep[see e.g.][and references therein]{Sefusatti_etal_2009,Biagetti:2013sr,Becker:2012je,Oppizzi:2017nfy}. However, the obvious drawback of such a simple scenario lies in the divergence of the $f_{\rm NL}(k)$ function at small scales, with a formally infinite amount of non-Gaussianity for $k\rightarrow \infty$.
This would, for example, induce a correlation between CMB spectral distortion and primary anisotropies, which, for large enough values of $|f_{\rm NL, 0}|$ and $n_{\rm NG}$, would have been detected in the \textit{Planck} and FIRAS data \citep{Rotti:2022lvy, Bianchini:2022dqh}.

To avoid such {\em ``small-scale catastrophe''}, in the present work we will consider a different functional form for $f_{\rm NL}^{\rm loc}(k)$ that follows the same evolution of Eq.~\ref{eq:runningfNL} around the pivot scale, but saturates to a maximum $|f_{\rm NL, max}^{\rm loc}|$ value at smaller scales, keeping such constant value deep into the $k\rightarrow \infty $ limit. This is realized through a hyperbolic tangent function by defining the running $f_{\rm NL}^{\rm loc}(k)$ according to:
\begin{equation}
f_{\rm NL}^{\rm loc}(k) = f_{\rm NL, max}^{\rm loc}\times \tanh \left( \alpha  \frac{k}{k_{0}}\right)^{n_{\rm NG}}
\label{eq:hyperbolicfNL}
\end{equation}
where the parameter $\alpha $ is defined as
\begin{equation}
\alpha \equiv \frac{1}{2} \ln \left( \frac{1+R^{1/n_{\rm NG}}}{1-R^{1/n_{\rm NG}}} \right)
\end{equation}
with $R\equiv f_{\rm NL, 0}^{\rm loc}/f_{\rm NL, max}^{\rm loc}$ and ensures that the resulting function has exactly the value $f_{\rm NL, 0}^{\rm loc}$ at the pivot scale and the desired maximum value of $|f_{\rm NL, max}^{\rm loc}|$.
As $f_{\rm NL, max}^{\rm loc}$ and $f_{\rm NL, 0}^{\rm loc}$ can be independently chosen to be  positive or negative, the parametrization of Eq.~\ref{eq:hyperbolicfNL} allows for both positive and negative non-Gaussianity. A similar idea of a small-scale saturation of $f_{\rm NL}$ has been recently proposed by \cite{Stahl:2024stz} although with a different shape of the $f_{\rm NL}(k)$ function.

In Fig.~\ref{fig:models}, the blue and magenta curves in the {\em top} panel show the evolution of two realizations of Eq.~\ref{eq:hyperbolicfNL} having the same value of $f_{\rm NL, 0}^{\rm loc}=10$ and the same large-scale running index $n_{\rm NG}=2$, but different values of $f_{\rm NL, max}^{\rm loc}$ of $10^{3}$ and $10^{4}$, respectively. Similarly, the red and green curves in the {\em bottom} panel show the evolution for two models sharing the same $f_{\rm NL, 0}^{\rm loc}=-10$ and $n_{\rm NG}=2$, but saturating to the values of $f_{\rm NL, max}^{\rm loc}$ of $-10^{3}$ and $-10^{4}$, respectively.
These are the models that we 
investigate with simulations in Sections~\ref{sec:simulations} and \ref{sec:results}, and that are summarised in Table~\ref{tab:models}. As all these models share the same $\Lambda $CDM cosmology, their background expansion history %will be
is identical and their different evolution of linear and non-linear structure formation %will be
is determined only by the different statistical properties of the initial density field dictated by their respective non-Gaussianity parameterized according to Eq.~\ref{eq:hyperbolicfNL}.
\begin{table}
\begin{center}
\begin{tabular}{lccc}
\hline
\hline
Model & $f_{\rm NL, 0}^{\rm loc}$ & $f_{\rm NL, max}^{\rm loc}$ & $n_{\rm NG}$ \\ 
\hline
\hline
$\Lambda $CDM-Gaussian & $0$ & $0$ & $0$\\
NG+1e3 & $10$ & $1000$ & $2$\\
NG--1e3 & $-10$ & $-1000$ & $2$\\
NG+1e4 & $10$ & $10000$ & $2$\\
NG--1e4 & $-10$ & $-10000$ & $2$\\
\hline
\hline
\end{tabular}
\end{center}
\caption{The characteristic parameters of the cosmological models investigated in the present work.}
\label{tab:models}
\end{table}

\section{The Simulations}
\label{sec:simulations}

In the present section we describe the numerical setup for our N-body simulations and for the generation of their initial conditions, as well as the procedures to extract halos and voids catalogs from the simulations outputs.

\subsection{Numerical Setup}
\label{setup}

We have performed a suite of cosmological N-body simulations of a periodic box of $100\, h^{-1}$Mpc on a side, filled with $512^3$ particles, for each of the models described above and summarised in Table~\ref{tab:models}, by means of the Tree-PM code {\small GADGET3} \citep[][]{gadget-2}. The resulting particle mass is $m=6.5\times 10^{8}\, h^{-1}$M$_{\odot}$ and the gravitational softening was set to $5\, h^{-1}$kpc, corresponding to about $1/40$-th of the mean inter-particle separation. All the simulations share the same standard cosmological parameters as summarised in Table~\ref{tab:params}, consistent with the 2015 constraints from the {\em Planck} satellite mission \citep[][]{Planck_2015_XIII}, but have different statistics of the initial density field imprinted on the respective initial conditions (as described below). 

\subsection{Initial Conditions}
\label{ics}

We %have 
set up initial conditions with {\em local} and {\em scale-dependent} non-Gaussianity for our N-body simulations by means of the {\small PNGRUN} initial conditions generator code by \cite{Wagner_Verde_Boubekeur_2010}. The latter is based on the computation of the non-Gaussian contribution $\Phi ^{\rm NG}$ to the gravitational potential $\Phi = \Phi ^{\rm G}+\Phi ^{\rm NG}$ starting from the desired {\em bispectrum} $B(k_{1},k_{2},k_{3})$ of the potential field, which is defined in Fourier space as:
\begin{equation}
\langle \Phi _{k_{1}}\Phi _{k_{2}}\Phi _{k_{3}}\rangle = (2\pi )^{3}\delta ^{D}(\vec{k}_{1}+\vec{k}_{2}+\vec{k}_{3})B(k_{1},k_{2},k_{3})\,
\end{equation}
where $\Phi _{k}$ is the Fourier transform of the real-space potential $\Phi $.

For the
{\em local} shape (i.e. dominated by the squeezed configuration $k_1 \ll k_2\sim k_3$) the bispectrum takes the form:
\begin{equation}
\label{eq:phi_bispectrum}
B(k_{1}, k_{2}, k_{3}) = 2f_{\rm NL}^{\rm loc}(k)\left[ P_{1}P_{2} + P_{2}P_{3} + P_{1}P_{3}\right]
\end{equation}
where $P_{i}\equiv P(k_{i})$ is the power spectrum at the wave mode $k_{i}$ and where we have now assumed $f_{\rm NL}^{\rm loc}(k)$ to be scale-dependent, with the scale $k$ corresponding to the two long modes of the local configuration, i.e. $k\equiv k_{2}\sim k_{3}$.

The non-Gaussian component of the gravitational potential can then be computed as:
\begin{equation}
\label{eq:phi_ng}
\Phi _{\vec{k}}^{\rm NG} = \frac{1}{6}\sum_{\vec{k}'} B(k, k', |\vec{k}+\vec{k}'|)\frac{\Phi _{\vec{k}'}^{\rm *G}}{P(k')}\frac{\Phi _{\vec{k}+\vec{k}'}^{\rm G}}{P(|\vec{k}+\vec{k}'|)}\,
\end{equation}
where $\Phi _{\vec{k}}^{\rm G}$ is a random realization of a Gaussian field with the power spectrum given by $P(k)\propto k^{n_{s}-4}$ and $n_{s}$ is the scalar spectral index.

Once the Fourier-space non-Gaussian potential $\Phi _{\vec{k}}^{\rm NG}$ has been computed, the linear density field $\delta _{\vec{k}}$ at redshift $z$ is derived from the potential $\Phi _{\vec{k}} = \Phi _{\vec{k}}^{\rm G} + \Phi _{\vec{k}}^{\rm NG}$ through the transfer function $T(k)$ and the Poisson equation:
\begin{equation}\label{eq:delta_phi}
\delta _{\vec{k}}(z) = \frac{2}{3}\frac{k^{2}T(k)D(z)}{\Omega _{\rm M}H_{0}^{2}}\Phi _{\vec{k}}
\end{equation}
where $D(z)$ is the linear growth factor, $\Omega _{\rm M}$ is the present-day matter fraction and $H_{0}$ is the Hubble constant. With this density field it is then possible to displace N-body particles from a regular cartesian grid according to second order Lagrangian perturbation theory (2LPT) at the desired starting redshift of the simulations, which for our investigation is chosen to be $z_{i}=49$.

The statistical properties of the initial particle distribution are inherited from the primordial gravitational potential ones. In particular, the power spectrum $P_{\Phi}(k)$ of the potential $\Phi = \Phi^{\rm G} + \Phi^{\rm NG}$ can be written as
\begin{equation}
\label{eq:pk_phi}
    P_{\Phi}(k)=P(k) + P^{\rm NG}(k),
\end{equation}
where $P(k)$ and $P^{\rm NG}(k)$ are the power spectra of $\Phi^{\rm G}_{\vec{k}}$ and $\Phi^{\rm NG}_{\vec{k}}$, respectively. Following from the definition of $\Phi^{\rm NG}_{\vec{k}}$ in Eq. \eqref{eq:phi_ng}, we note here that the cross-term $2\langle\Phi^{\rm G}_{\vec{k}}\Phi^{\rm NG}_{\vec{k}}\rangle$ does not contribute to $P_{\Phi}(k)$ as it involves an odd number of Gaussian fields.

On the other hand, the non-Gaussian component power spectrum $P^{\rm NG}(k)$ scales as $|f_{\rm NL, max}^{\rm loc}|^2$. Consequently, models with opposite non-Gaussianity parameters share the same primordial potential power spectrum; however, they have opposite bispectrum, according to Eq. \eqref{eq:phi_bispectrum}. {We point out that Eq. \eqref{eq:localNG} is expressed in terms of the \emph{primordial} potential.
Rewriting it in terms of density perturbation via the Poisson equation, within linear evolution, one can factor out the overall $D^{-1}(z)$ redshift scaling. Then, it becomes clear that the term proportional to $f_{\rm NL}^{\rm loc}$, quadratic in the density perturbations, is suppressed by an extra $D^{-1}(z)$ with respect to the Gaussian one \citep{Dalal_etal_2008}.
The effect of this scaling will be most apparent in the matter power spectrum, which we discuss in Section~\ref{sec:matter_power_spectrum}. For large and negative values of the non-Gaussian parameter, the terms linear and quadratic in $f_{\rm NL}^{\rm loc}$ are competing with each other at high redshift, while at lower redshift the quadratic terms become subdominant. This behaviour has been studied on mildly non-linear scales and corresponds to the different redshift dependence of the one-loop contributions in \cite{Taruya:2008pg}.}

\subsection{Halo identification}
\label{halofind}

To obtain a catalog of halos in each cosmological scenario, we post-process the output snapshots of all our simulations to identify particle groups through a Friends-of-Friends algorithm \citep[FoF,][]{Davis_etal_1985}, adopting a linking length of $20\%$ the mean inter-particle separation. The resulting groups composed by at least 32 particles are   stored into a catalog of FoF halos.
We then run the {\small SUBFIND} algorithm \citep[][]{Springel_etal_2001} on the FoF catalog to identify gravitationally bound substructures, and we store the resulting halos having at least 20 gravitationally bound particles, for which we compute spherical overdensity properties as e.g. the virial mass $M_{200}$ and virial radius $R_{200}$, defined by the relation:
\begin{equation}
\frac{4\pi }{3}R_{200}^{3}\Delta \rho_{\rm crit} = M_{200}
\end{equation}
where $\rho_{\rm crit}\equiv 3H^{2}/8\pi G$ is the critical density of the universe and $\Delta = 200$.

We only consider halos with $M_{200}$ in the range $\left[ 10^{11}\,, 10^{14}\right] $ M$_{\odot }/h$ since lower mass halos are too poorly resolved and higher mass halos are too rare to provide robust statistical conclusions. 

\subsection{Void identification}
\label{voidfind}

Besides detecting halos in the simulated volumes of our simulations, we also identify cosmic voids -- i.e.  regions of space with a lower density compared to the average density of the box -- by means of the publicly available void finder {\small VIDE} \citep[][]{VIDE} based on the {\small ZOBOV}  algorithm \citep{Neyrinck_2008}, which 
identifies voids in any density field sampled by a discrete set of particles. The algorithm exploits a Voronoi tessellation scheme and identifies local density minima by selecting Voronoi cells surrounded only by cells with a lower Voronoi volume \citep[see][for a detailed description of {\small ZOBOV}]{Neyrinck_2008}. The Voronoi cell with the largest volume in each void then identifies the particle at the local density minimum, whose position is assumed as the center of the void for the computation of void density profiles described below in Section~\ref{sec:void_profiles}.

The resulting catalog of underdense regions of space is then processed by the {\small VIDE} toolkit by performing various possible selections of the void sample as e.g. different cuts on the void density contrast or on the void central overdensity. This allows us to select void samples with defined properties or specific thresholds on the statistical significance of the void itself. 

In particular, for the analysis discussed in this work we 
only consider voids in the distribution of dark matter particles having a central density below $0.2$ times the mean density of the universe and with a density contrast between the most underdense particle of the void and the void boundary (i.e. the radius at which the void radial density profile starts decreasing towards the cosmic average density) larger than $1.57$ \citep[which corresponds to a probability that the void arises as a result of Poisson noise below $\sim5\%$, see again][]{Neyrinck_2008}. 

A standard quantity to characterize cosmic voids and study their relative abundance
is given by their effective radius $R_{\rm eff}$, defined
from the Voronoi volume of the void as the radius of a sphere having the same volume as the void:
\begin{equation}
V_{\rm VOID} \equiv {\sum\limits_{i=1}^N V^{p}_i }= \frac{4}{3} \pi R_{\rm eff}^3 \,.
\end{equation}
We 
compute this quantity for all the voids fulfilling our selection criteria and 
describe the void radial density profiles in units of this effective radius below (see Section~\ref{sec:void_profiles}).

\section{Results}
\label{sec:results}

In this section, we present the main outcomes of our numerical investigation and discuss their main implications in terms of possible observational constraints of the proposed scale-dependent non-Gaussianity models and in terms of possible degeneracies with other extensions of the standard cosmological scenario.

\subsection{Matter distribution}

\begin{figure*}
\includegraphics[width=0.32\textwidth]{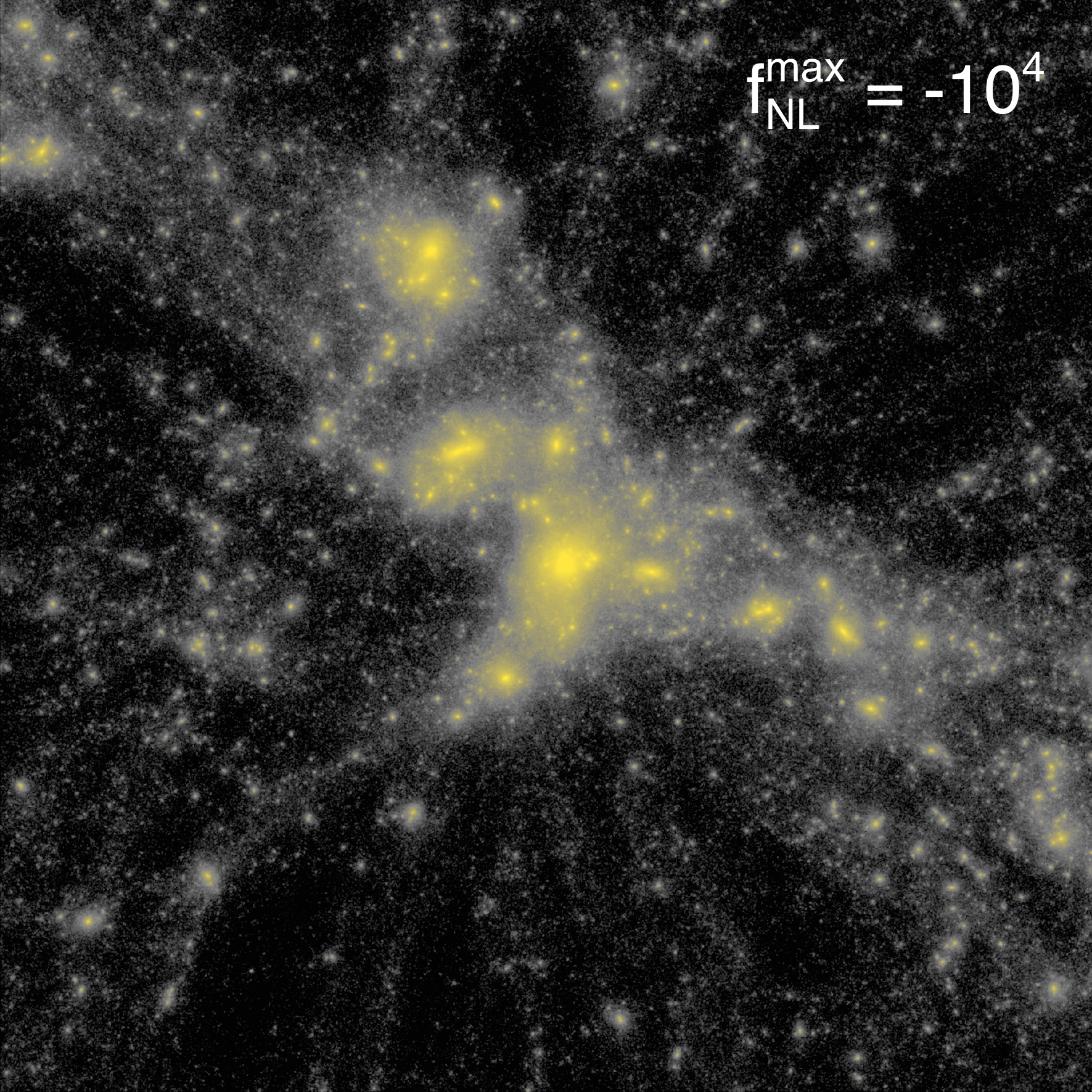}
\includegraphics[width=0.32\textwidth]{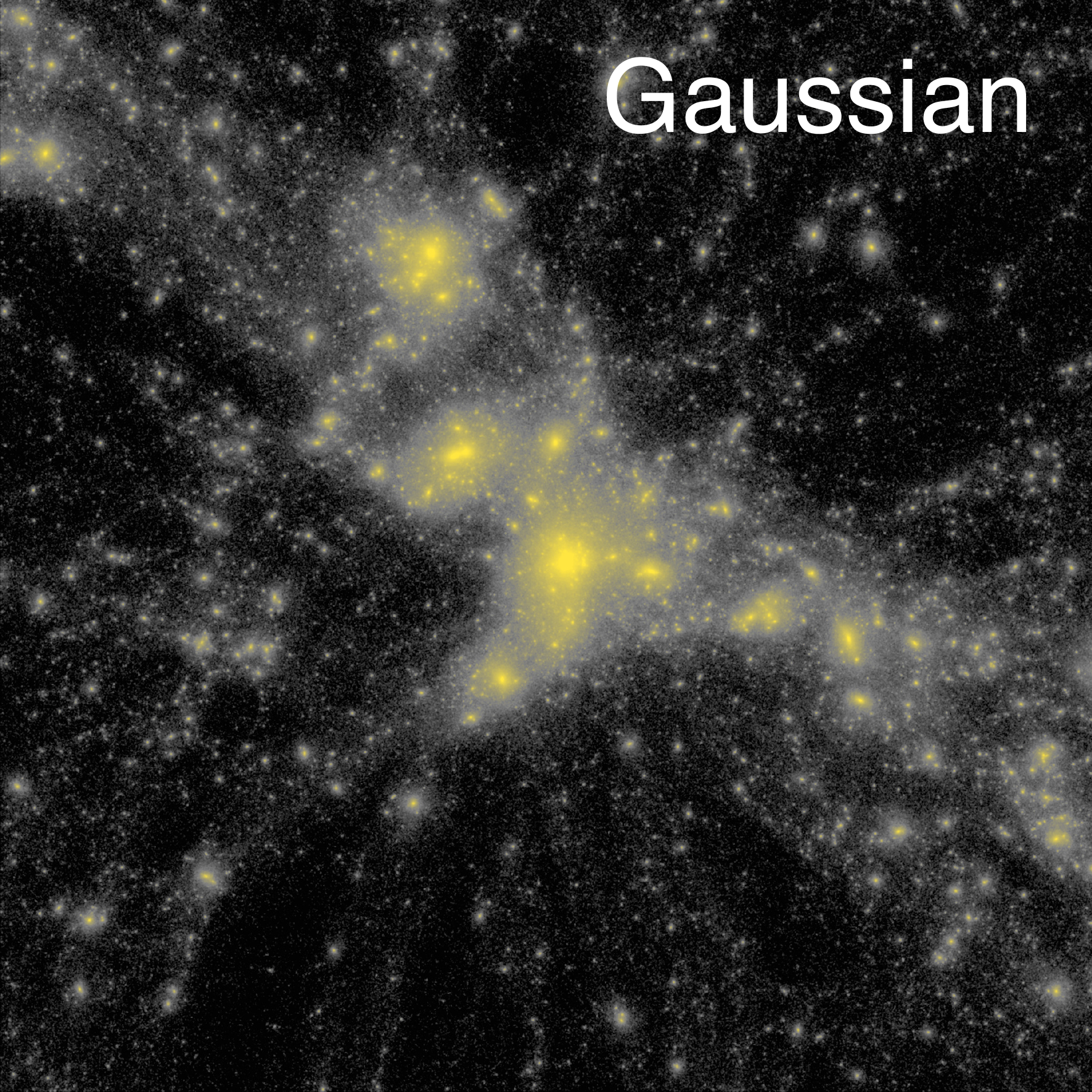}
\includegraphics[width=0.32\textwidth]{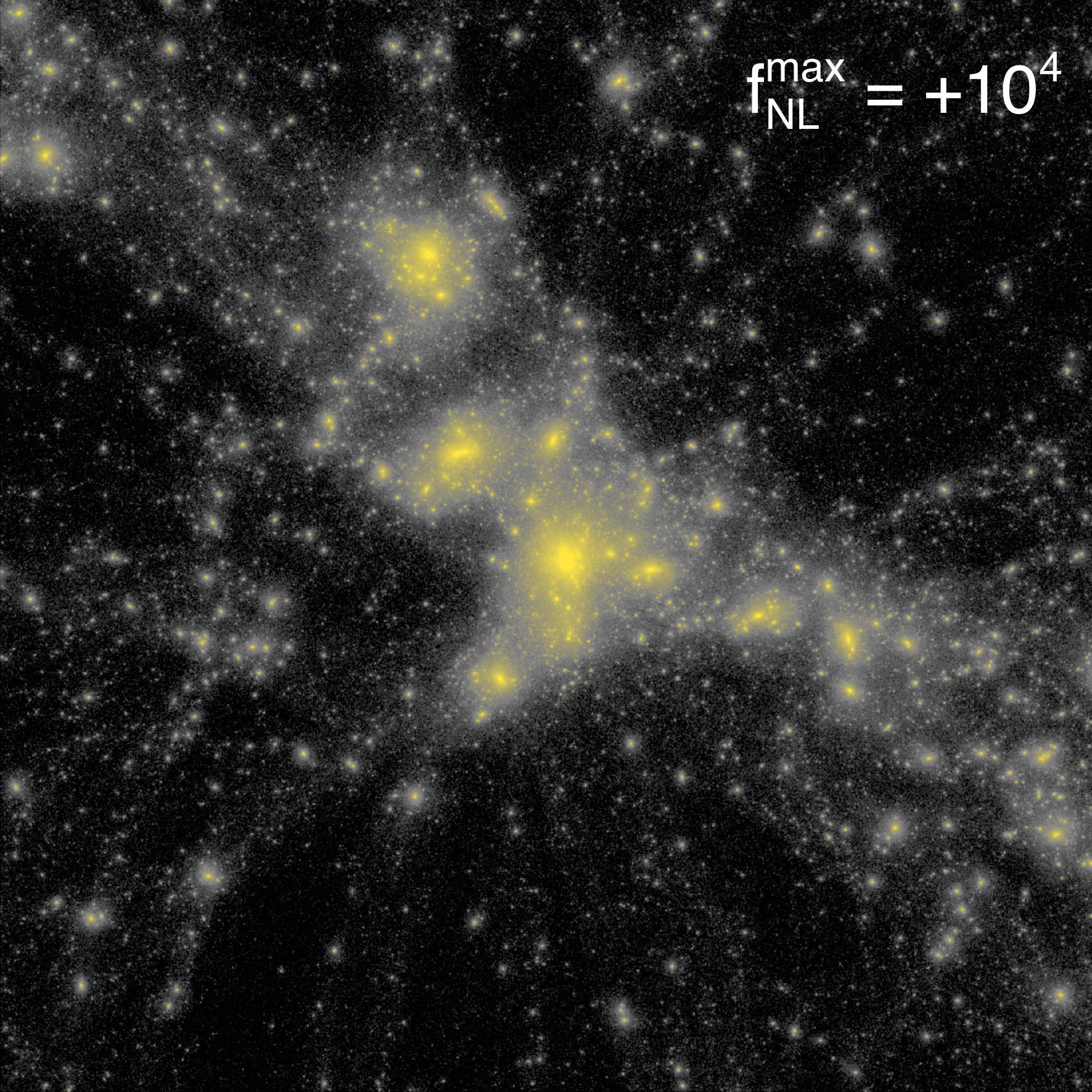}
\caption{The projected density distribution in a $25$ Mpc$/h$ box centered on the most massive halo identified in the Gaussian simulation ({\em center}) and in the two most extreme models with $f_{\rm NL, max}^{\rm loc}=-10^{4}$ ({\em left}) and $f_{\rm NL, max}^{\rm loc}=+10^{4}$ ({\em right}).}
\label{fig:slices}
\end{figure*}

In Fig.~\ref{fig:slices} we show the density field around the most massive halo (with mass $M_{200c}\approx 1.3\times 10^{15} $ M$_{\odot }/h$) extracted from the $z=0$ snapshot of the reference Gaussian $\Lambda $CDM simulation (central panel) and  the corresponding region for the two most extreme non-Gaussian realizations
with negative and positive values of the $f_{\rm NL}^{\rm loc}(k)$ function (left and right panels, respectively). The images show the projected mass distribution in a cube of $25\, h^{-1}$Mpc per side. 

As one can see from the figures, the shape of the large-scale matter distribution is the same in all the simulations, as a consequence of assuming identical phases for the random realization of the power spectrum in the initial conditions, as discussed in Section~\ref{ics} above. Nonetheless, some differences appear in the location of individual substructures as well as in the concentration of the density peaks, even though no clear trend can be identified by eye in these minimal changes from one simulation to another. 

This result already shows -- qualitatively -- that even the most extreme scenarios considered in our work, with a maximum value of $|f_{\rm NL}^{\rm loc}|=10^{4}$ at small scales, do not significantly change the overall large-scale matter distribution, thanks to the steep suppression of $|f_{\rm NL}^{\rm loc}|$ at progressively larger scales as shown in Fig.~\ref{fig:models}. However, as we will describe below, the statistical analysis of the matter distribution and the halo and void structural properties will highlight differences  among the models  that become important at small scales and  can be tested by present and future observations.

\subsection{Matter power spectrum}
\label{sec:matter_power_spectrum}

\begin{figure*}
\includegraphics[width=0.48\textwidth]{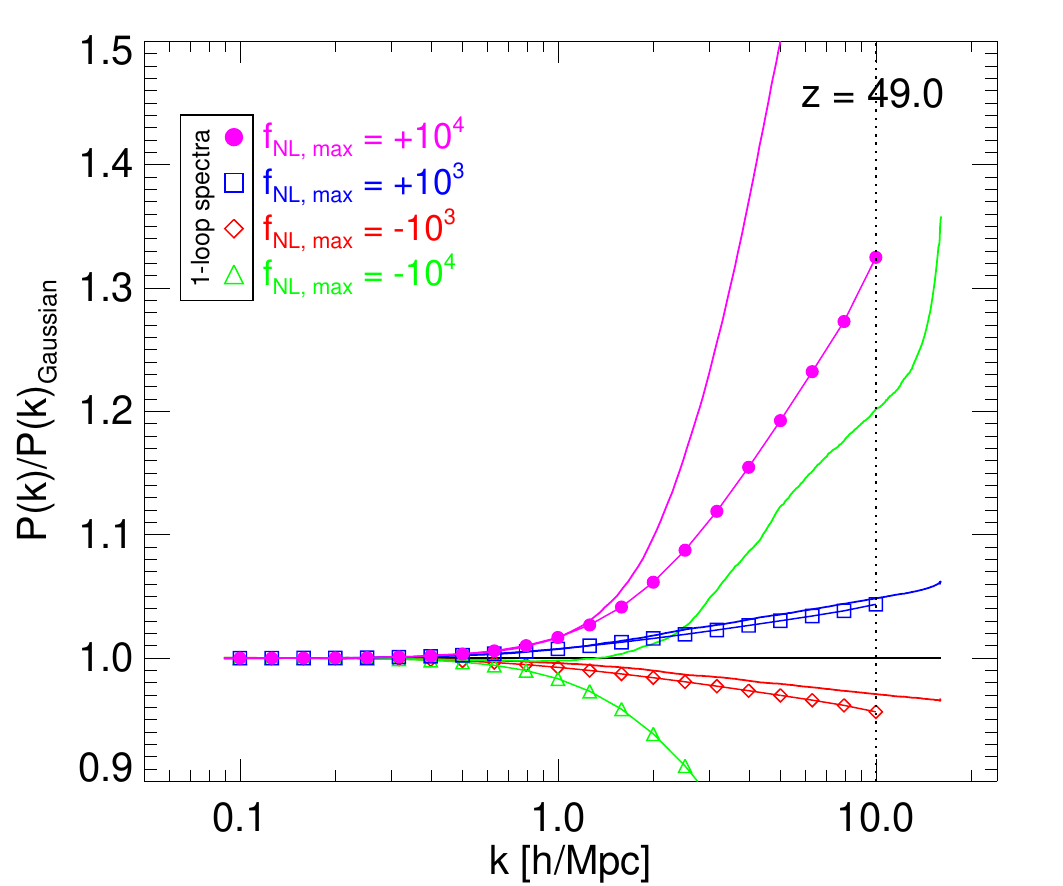}
\includegraphics[width=0.48\textwidth]{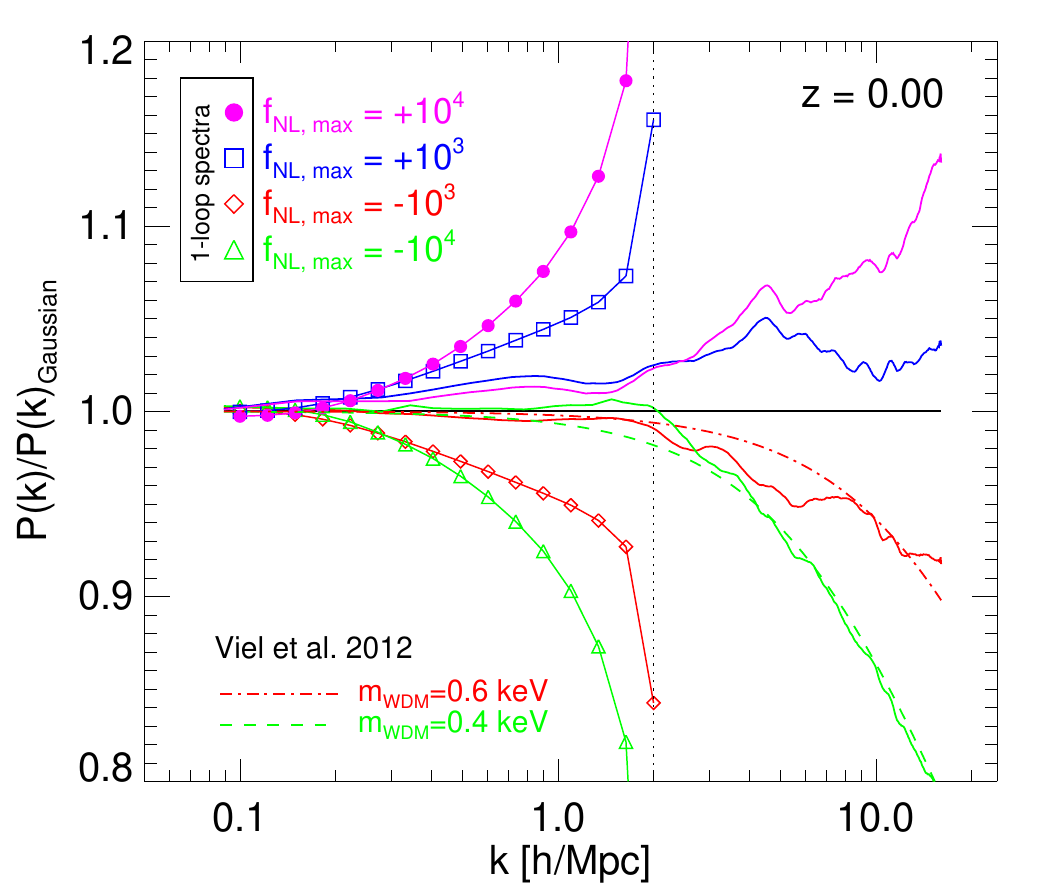}
\caption{{The matter power spectrum ratio to the standard Gaussian case at the initial conditions ($z=49$, left panel) and at the present epoch ($z=0$, right panel). The red dot-dashed and green dashed curves in the {\em right} plot represent the expected suppression in the matter power spectrum for a Warm Dark Matter model with $m_{\rm WDM}=0.6$ keV and $m_{\rm WDM}=0.4$ keV, respectively, as computed with the fitting formula of \cite{Viel_etal_2012}. {Lines with symbols indicate the 1-loop predictions for the power spectrum ratio computed as described in the text. The predictions have been truncated at the scale beyond which the 1-loop perturbative approach becomes ill-defined, which is indicated by the vertical dotted lines in the plots.}}  }
\label{fig:powerspec}
\end{figure*}

We start our analysis by measuring the non-linear matter power spectrum from the simulations and comparing the non-Gaussian with the Gaussian ones.
In Fig.~\ref{fig:powerspec} we show this ratio for the initial conditions of the simulations at $z_{i}=49$ (left panel) and for the final $z=0$ snapshot (right panel). 

As expected, at  $k < 0.2\, h/$Mpc all non-Gaussian cases are indistinguishable from the Gaussian one: the differences  appearing at small scales however may  sometimes appear counter-intuitive.

In particular, the {\em left} plot clearly shows how, for the largest values of the maximum non-Gaussianity ($\pm 10^{4}$), the contribution $P^{\rm NG}(k)$ in Eq. \eqref{eq:pk_phi}, scaling as $|f_{\rm NL, max}^{\rm loc}|^2$, can significantly increase the resulting power at small scales -- regardless of the sign of the $f_{\rm NL}^{\rm loc}(k)$ function --. 
For smaller $f_{\rm NL}(k)$ values, instead, this contribution is subdominant with respect to the one that is linear in $f_{\rm NL, max}^{\rm loc}$ (which vanishes in $P_{\Phi}(k)$, but is present at the particle distribution level, due to 2LPT).
Consequently, the deviations from the Gaussian case are almost symmetrical for positive and negative scenarios with $|f_{\rm NL, max}^{\rm loc}|=10^3$.

The $z=0$ plot, instead, shows that a clear hierarchy of the models is restored in the late universe due to the gravitational evolution of the primordial density perturbations. {Thanks to the different redshift scaling illustrated in Section~\ref{ics}, terms which are linear in $f_{\rm NL}^{\rm loc}$ dominate other PNG contributions. Thus} in this case, models with positive (negative) $f_{\rm NL}^{\rm loc}(k)$ show a small-scales  enhancement (suppression)
compared to the Gaussian model. 
In particular, for any given sign of the $f_{\rm NL}^{\rm loc}(k)$ function, the shape of the deviation in the matter power spectrum is identical for different values of $f_{\rm NL, max}^{\rm loc}$ up to a scale of $\approx 5\, h/$Mpc. Beyond this scale, the deviation for the models with the largest and smallest $f_{\rm NL, max}^{\rm loc}$ start to deviate from each other, with the former models showing a steeper evolution resulting in progressively larger deviations at smaller and smaller scales. This reflects the fact that the shape of the $f_{\rm NL}^{\rm loc}(k)$ function is the same for all models at large scales -- corresponding to a power-law with slope $n_{\rm NG}=2$ -- while the hyperbolic term of Eq.~\ref{eq:hyperbolicfNL} giving rise to the saturation of $f_{\rm NL}^{\rm loc}(k)$ to a given $f_{\rm NL, max}^{\rm loc}$ kicks in only at smaller scales.

{We also plot for comparison the 1-loop predictions for the power spectrum ratio computed within the framework of the effective field theory (EFT) \cite{Baumann:2010tm,Carrasco:2012cv} (see \cite{Cabass:2022avo,Ivanov:2022mrd} for recent reviews) in the presence of the non-Gaussian initial conditions \footnote{Note that there is only one $\mathcal{O}(f_{\rm NL})$ non-Gaussian contribution to the 1-loop matter power spectrum, i.e. the $P_{12}$ term \cite{Taruya:2008pg,Assassi:2015jqa}, which is calculated here as $P_{12}(k;z)=2\int_{\vec{p}}\;f_{\rm NL}^{\rm loc}(p)\,F_2(\vec{p},\vec{k}-\vec{p})\;B(k,p,|\vec{k}-\vec{p}|;z)$, where $\int_{\vec{p}}=\int\frac{\rm{d}^3p}{(2\pi)^3}$ and $F_2$ is the symmetrized second order density kernel from Standard Perturbation Theory (see e.g. \cite{Bernardeau:2001qr}). The bispectrum in the integral is the shape of the linear propagated primordial bispectrum of Eq. \eqref{eq:phi_bispectrum}, i.e. $B(k_1,k_2,k_3;z)=2\,M(k_1;z)M(k_2;z)M(k_3;z)\,[P_1P_2+P_2P_3+P_1P_3]$, where the expresion for $M(k;z)$ (i.e. $\delta _{\vec{k}}(z) = M(k;z)\Phi _{\vec{k}}$) is shown in Eq. \eqref{eq:delta_phi}.} \cite{Assassi:2015jqa}, described in Sec. \ref{ics}. The results include IR resummation \cite{Senatore:2014via,Vasudevan:2019ewf}, as well as the higher
derivative terms (counterterms) required by the renormalization procedure. As one can see in the plots, the 1-loop EFT predictions recover the power spectrum at the initial condition snapshot (i.e. $z=49$) up to small scales and only for the ``NG+1e3" and ``NG-1e3" models (blue and red points in Fig.~\ref{fig:powerspec}). At these redshifts, the linear regime is significantly extended and with it the predictability of the EFT model. For the ``NG+1e4" and ``NG-1e4" models (magenta and green points in Fig.~\ref{fig:powerspec}), the perturbative description of the non-Gaussian part breaks down much earlier, due to their large $f_{\rm NL, max}^{\rm loc}$ values. The discrepancy of these PNG models with simulation results, can also be atrributed to the fact that only the non-Gaussian term linear in $f_{\rm NL}^{\rm loc}$ is considered in the 1-loop model, while the quadratic contribution can be important at these high redshifts (see Sec.~\ref{setup} for a discussion). At the present epoch (i.e. $z=0$), where structures have undergone non-linear evolution, the perturbative approach breaks down at much larger scales, failing to recover the power spectrum evolution for $k \gtrsim 0.2\, h/$Mpc. Thus, numerical simulations are necessary to obtain reliable predictions at non-linear scales for models with this type of scale dependence of the non-Gaussianity parameter $f_{\rm NL}$.}

% As one can see in the plots, the 1-loop predictions fail to recover the power spectrum evolution at small scales, showing that numerical simulations are necessary to obtain reliable predictions at non-linear scales for models with this type of scale dependence of the non-Gaussianity parameter $f_{\rm NL}$.

Interestingly, we also note that the models with negative $f_{\rm NL}^{\rm loc}(k)$ give rise to a suppression of the small-scale power with a shape similar to that of a thermal cutoff, that may resemble a Warm Dark Matter \citep[WDM, see e.g.][]{Bode_Ostriker_Turok_2001} density distribution, thereby giving rise to an observational degeneracy between the two independent phenomena of primordial non-Gaussianity and of a thermal suppression of the density perturbations at small scales due to free-streaming. As a reference, we have overplotted in Fig.~\ref{fig:powerspec} the expected suppression of non-linear power associated with a WDM particle candidate with mass $m_{\rm WDM}=0.4$ keV (green dashed curve) and $m_{\rm WDM}=0.7$ keV (red dot-dashed curve) using the fitting formula of \citep[][]{Viel_etal_2012}, { where we have used the values $\nu = 3$, $l = 0.5$, $s = 0.6$ for the fitting parameters}. As one can see from the figure, the shape and amplitude of the suppression very closely match the ones coming from our non-Gaussian scenarios with negative $f_{\rm NL}^{\rm loc}(k)$. Similar types of observational degeneracies between theoretically independent modifications of the standard cosmological model have been recently investigated in \cite{Baldi_etal_2014,Baldi_Villaescusa-Navarro_2018,Hashim:2018dek}. As we will see below, the degeneracy identified in the shape of the non-linear matter power spectrum between WDM scenarios and our proposed models of scale-dependent non-Gaussianity will  also appear in other observables (though not in all of them) investigated in the present work. This indicates the possible presence of 
a significant degeneracy between these two independent scenarios that should be further investigated.

\subsection{Halo Mass Function}

\begin{figure}
\center
\includegraphics[width=0.7\columnwidth]{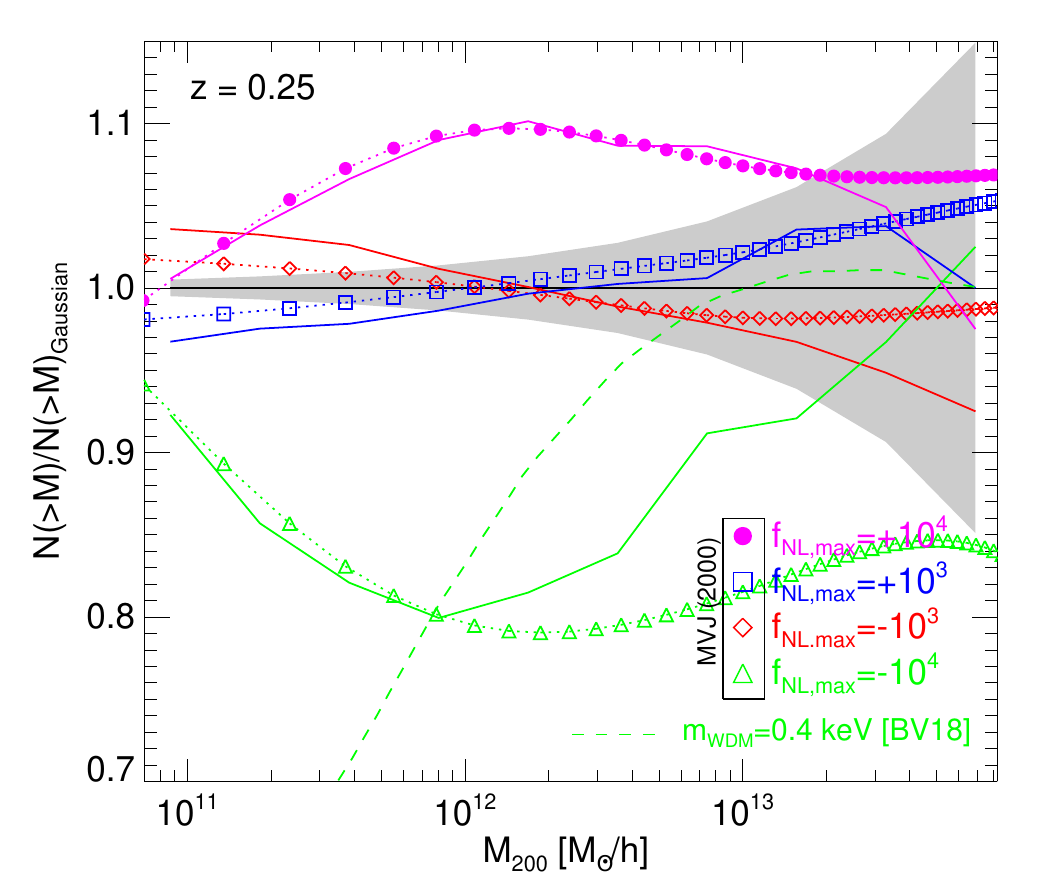}
\caption{{ The halo mass function ratio to the standard Gaussian case for the different models. {Dotted lines with symbols indicate the analytical predictions \eqref{eq:rng_pred} with $q=1.5$ for all the models except $f_{\rm NL,max}=-10^{4}$, for which $q=4$.} The dashed green curve shows the suppression obtained for a Warm Dark Matter model with $m_{\rm WDM}=0.4$ keV as obtained from the simulations performed in \cite{Baldi_Villaescusa-Navarro_2018}. As one can see, differently from the case of the non-linear matter power spectrum depicted in Fig.~\ref{fig:powerspec}, the Warm Dark Matter suppression shows a completely different shape with respect to the non-Gaussian cases.}}
\label{fig:HMF}
\end{figure}

With the halo catalogs compiled as described in Section~\ref{halofind}, we
computed the cumulative halo mass function -- i.e. the abundance of halos as a function of their virial mass $M_{200c}$ -- for all the cosmological models under investigation. The resulting cumulative mass functions {at $z=0.25$\footnote{In all our plots we choose to show results at $z=0.25$ rather than at $z=0$ as this is the redshift for which we can access comparison data for the behaviour of the degenerate WDM model as obtained by \cite{Baldi_Villaescusa-Navarro_2018}}} are shown in Fig.~\ref{fig:HMF}, normalized to the standard Gaussian reference case, {where we have binned  our halo catalog into $10$ logarithmically equispaced mass bins covering our selected mass range  $[10^{11},10^{14}] $ M$_{\odot}/h$}. By looking at the plot, one can immediately notice how the deviation from the reference scenario is symmetric for positive and negative values of the non-Gaussianity function for the models with the lowest value of $|f_{\rm NL, max}^{\rm loc}|=10^{3}$, while this symmetry is partly lost for the models with the largest value of $|f_{\rm NL, max}^{\rm loc}|=10^{4}$. 

In particular, the former models show a modest suppression (enhancement) of the halo abundance at the low-mass end of our halo sample for positive (negative) non-Gaussianity and a corresponding enhancement (suppression) of comparable magnitude at the high-mass end of the sample. The transition between these two regimes occurs at a mass of {$\approx 1.7\times 10^{12}\, h^{-1}$M$_{\odot}$}, where the two curves cross each other and simultaneously cross the reference Gaussian case. The overall relative deviation from the standard model never exceeds $3\%$ over the whole sampled mass range. 

The situation is significantly different for the latter models, which show a much more significant and less symmetric deviation from the Gaussian case. More specifically, the model with positive non-Gaussianity shows an enhancement of the abundance of halos over the whole mass range covered by our sample, with a deviation of $2-4\%$ at the extremes of the sample and a maximum deviation of $\approx 8\%$ at intermediate masses (interestingly, the peak of the deviation is reached for the same mass bin where the transition between enhancement and suppression occurs for the lower $|f_{\rm NL, max}^{\rm loc}|$ models, {$\approx 1.7\times 10^{12}\, h^{-1}$M$_{\odot}$}). On the contrary, the model with negative non-Gaussianity shows a suppression of the halo abundance over the whole mass range, recovering the expected abundance of the standard Gaussian case only at the largest mass bin of the sample. Also in this case, the deviation from the reference cosmology is weaker at the two extremes of the sample ($\approx 7.5\%$ at the low-mass end, and negligible -- as just mentioned -- at the high-mass end), and reaches a maximum of $\approx 20\%$ at intermediate masses. In this case, however, the maximum deviation does not occur at the same mass as for the positive non-Gaussian case, with a peak of the suppression at $\approx 8\times 10^{11}\,h^{-1}$M$_{\odot }$. Again, this phenomenology reflects the stronger impact of higher-order terms on the density distribution for models with a larger value of $|f_{\rm NL, max}^{\rm loc}|$.
{In Fig.~\ref{fig:HMF} we also plot, as dotted lines with symbols, the prediction of the mass function ratio $R^{NG}(M,z)$ obtained for the proposed models following the approach detailed in \cite{Matarrese_Verde_Jimenez_2000}.}
{More specifically, $R^{NG}(M,z)$ can be derived in the framework of the Press-Schechter formalism by exploiting the saddle-point technique, obtaining the following analytical expression:
\begin{equation}\label{eq:rng_pred}
R^{NG}(M,z)=\exp\left(\frac{\Delta_c^3(z) S_{3,M}}{6\sigma^2_M}\right)\left|\frac{\Delta_c(z)}{6\sqrt{1-\Delta_c(z)S_{3,M}/3}}\frac{{\rm d}S_{3,M}}{{\rm d}\ln \sigma_M}+\sqrt{1-\Delta_c(z)S_{3,M}/3}\right| \,,
\end{equation}
where $\sigma_M^2$ and $S_{3,M}=\langle \delta_{M}^3\rangle/\sigma_M^4$ are respectively the variance and the normalized skewness of $\delta_M$, the linear density field smoothed on a mass scale $M$ at $z=0$. 
This expression depends on redshift only through the collapse threshold, $\Delta_c(z)=\sqrt{q}\delta_c D(z=0)/D(z)$, where $q$ is a fudge factor which varies with the details of the simulated halos, such as the halo finder algorithm considered \cite{Grossi_etal_2007,wagner2012}.

A choice of $q=1.5$ maximizes the accuracy of the predictions for most of the models considered here, except $f_{\rm NL,max}=-10^{4}$. For the latter one, we report in Fig.~\ref{fig:HMF} the analytical prediction with $q=4$.}

To continue our comparison of the effects of our models with negative non-Gaussianity to WDM cosmologies, we have overplotted as a reference in Fig.~\ref{fig:HMF} the mass function deviation from the standard $\Lambda $CDM cosmology for the case of a WDM particle candidate with $m_{\rm WDM}=0.4$ keV (i.e. corresponding to the green-dashed curve in the {\em right} panel of Fig.~\ref{fig:powerspec}) obtained from the  simulations of \cite{Baldi_Villaescusa-Navarro_2018} (BV18 hereafter). We stress here that the simulations of BV18 have the same specifications (box size and particle number) and cosmological parameters (as detailed in Table~\ref{tab:params}) as our simulations, but different statistical realisations (i.e. a different initial conditions random phase). As one can see from the comparison with the solid green curve, differently from what was found for the nonlinear matter power spectrum, the impact of the two scenarios on the halo mass function is starkly different. Although both models result in a suppression of the halo abundance at small masses and recover the standard $\Lambda $CDM prediction at the largest masses of the sample, the shape and overall amplitude of the effect are completely different in the two cases, with WDM showing a much flatter behavior at large masses and a much steeper suppression for masses below $10^{13}\,h^{-1}$M$_{\odot }$. This result shows that, despite very similar 2-point statistics, the two models are significantly different at the level of fully nonlinear collapse of structures. Having in mind the halo model approach, this result suggests that also halo concentrations should behave differently in the two scenarios, as we will indeed find below (see Section~\ref{sec:concentrations}).

\subsection{Halo Bias}

\begin{figure}
\center
\includegraphics[width=0.7\columnwidth]{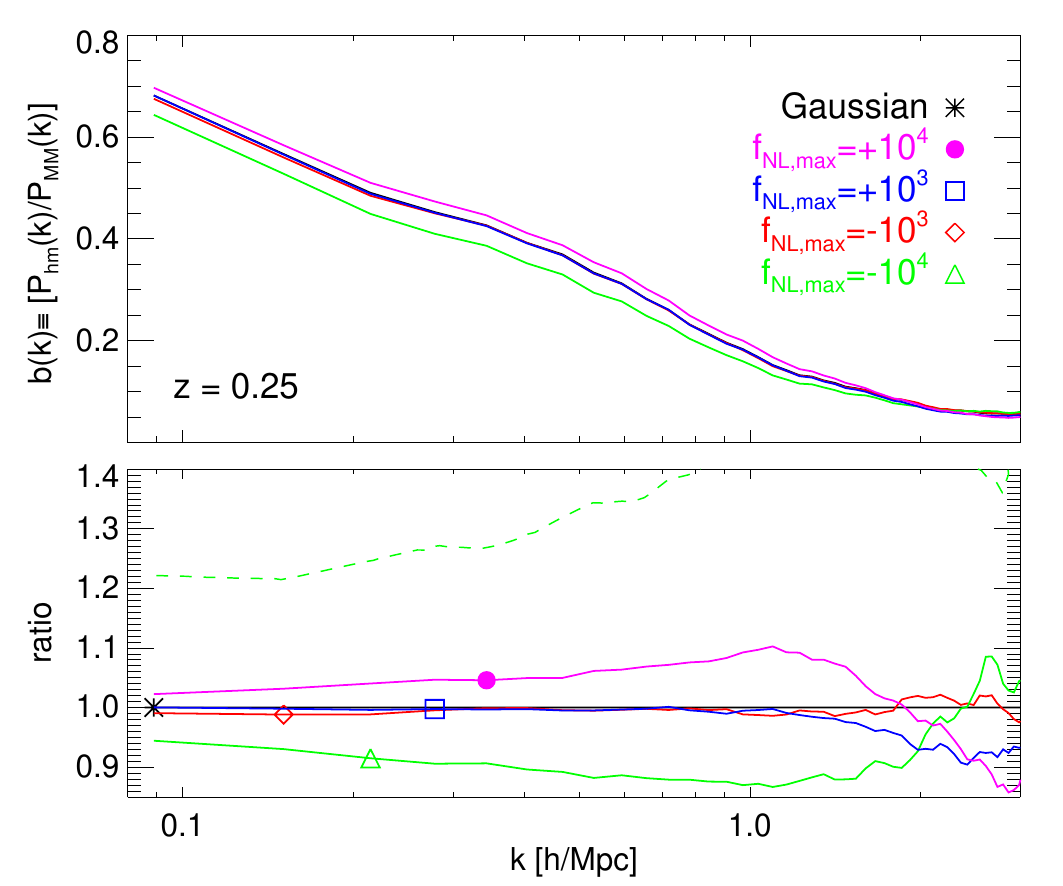}
\caption{{ The halo bias as a function of scale (top) and its ratio to the standard Gaussian case (bottom) for all simulations at $z=0.25$. The green dashed curve shows the bias ratio for a WDM model with $m_{\rm WDM} = 0.4$ keV as obtained by BV18}}
\label{fig:bias}
\end{figure}

In Fig.~\ref{fig:bias} we display the halo bias at $z=0.25$ as a function of scale computed as {the ratio between the  halo-matter cross power spectrum $P_{hm}(k)$ and the  matter auto power spectrum $P(k)$:
\begin{equation}
b(k)\equiv \frac{P_{hm}(k)}{P(k)}.
\end{equation}
}
{As for previous observables, we show the ratio of the bias to the standard Gaussian case, for the four non-Gaussian scenarios under investigation. As a general trend, we notice that the models with the lowest $|f_{\rm NL,max}|= 10^{3}$ show basically no deviations from the Gaussian bias except at the smallest available scales (beyond $1\, h$Mpc$^{-1}$) where the ratio becomes scale-dependent for the $f_{\rm NL,max}=+10^{3}$ model. The models with the largest $|f_{\rm NL,max}|=10^{4}$ instead show clear deviations from the Gaussian case, with the positive $f_{\rm NL}$ model showing a higher bias and the negative $f_{\rm NL}$ model showing a lower bias, again with the exception of the smallest scales where the ratio becomes scale dependent and the trend is inverted.

Unfortunately, the size of our simulations is too small to probe the linear bias, where scale-dependent features may appear again for all models in a range of scales that could be more easily probed by large-scale structure surveys. We will extend our analysis to larger cosmological volumes that will allow us to explore the linear bias regime in future work. Furthermore, differently from the case of the halo mass function, in this case the theoretical predictions obtained following \cite{Matarrese_Verde_Jimenez_2000} do not seem to accurately reproduce the bias ratio measured from the simulations, possibly due to the smaller scales and lower halo masses sampled in our work compared to the ones considered in \cite{Matarrese_Verde_Jimenez_2000}. We also defer a more detailed analysis of such comparison to future work.}

Finally, we compare again the bias ratio obtained for our non-Gaussian scenarios to the one observed for a WDM particle candidate with $m_{\rm WDM}=0.4$ keV in BV18, which is shown as a dashed green curve in the bottom panel of Fig.~\ref{fig:bias}. Evidently, as already shown for the halo mass function, the halo bias does not show the same degeneracy that was observed in the non-linear matter power spectrum (Fig.~\ref{fig:powerspec}), with the WDM model showing a substantial increase of the bias on all scales whereas its degenerate (in the power spectrum) non-Gaussian model $f_{\rm NL, max}=-10^{4}$ shows a suppression of the bias. 

\subsection{Concentrations-mass relation}
\label{sec:concentrations}

\begin{figure*}
\includegraphics[width=0.45\textwidth]{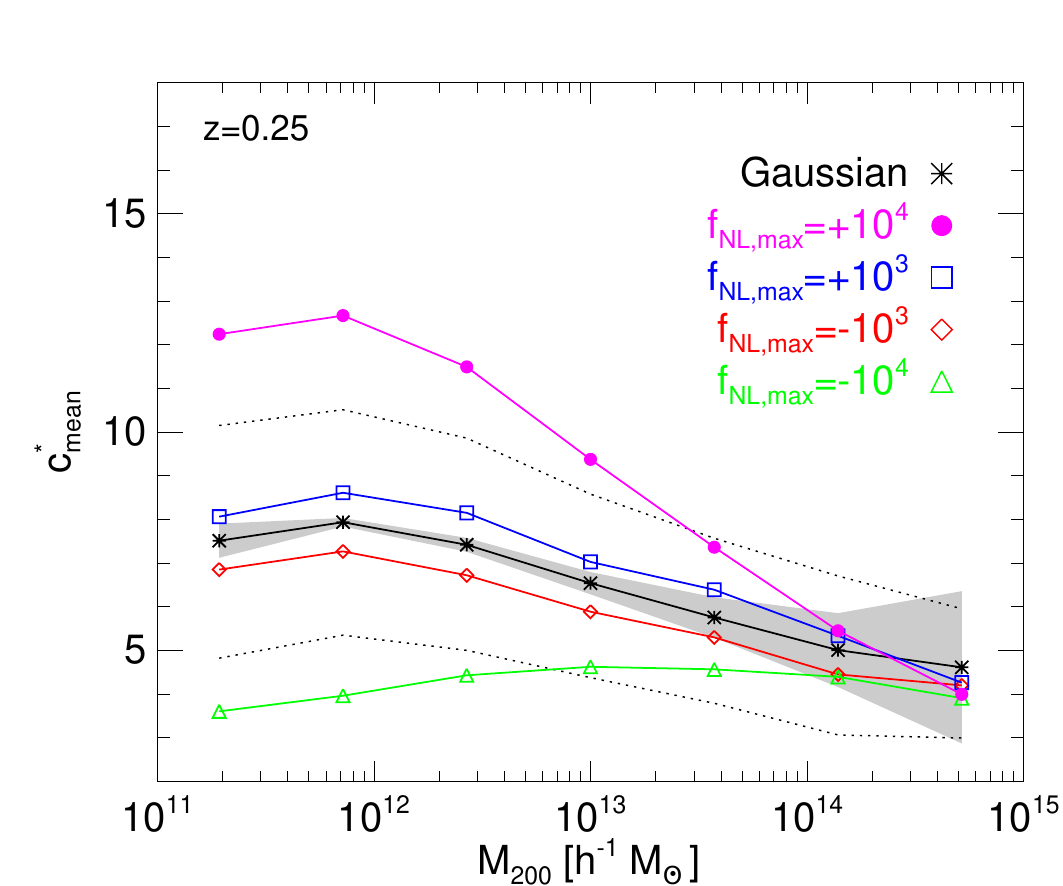}
\includegraphics[width=0.45\textwidth]{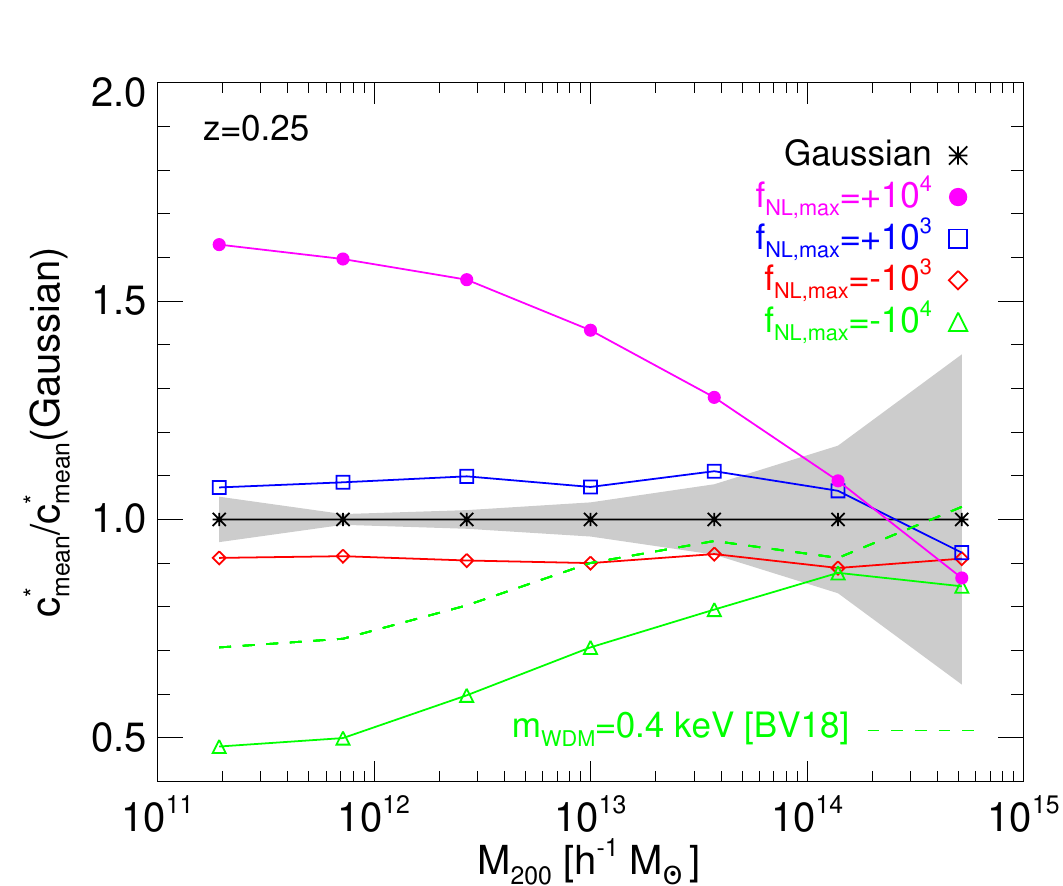}
\caption{{ The halo concentrations as a function of mass ({\em left}) and their ratio to the Gaussian case ({\em right}) for the different models. In the left plot, the dotted curves indicate the spread of $68\%$ of the halos in
each mass bin, while the grey-shaded area in both plots shows the Poissonian error associated with the number of halos in each bin. Also, in the {\em right} plot, the green dashed curve displays the suppression of halo concentrations for a Warm Dark Matter particle candidate with $m_{\rm WDM}=0.4$ keV, as obtained from the simulations of \cite{Baldi_Villaescusa-Navarro_2018}.}}
\label{fig:conc}
\end{figure*}
For all the halos in our sample we compute 
the concentration $c^{*}$ defined according to \cite{Aquarius} as:
\begin{equation}
\label{conc_vmax}
\frac{200}{3}\frac{c^{*3}}{\ln (1+c^{*}) - c^{*}/(1+c^{*})} = 7.213~\delta _{V}
\end{equation}
where $\delta _{V}$ is:
\begin{equation}
\delta _{V} = 2\left( \frac{V_{max}}{H_{0}r_{max}}\right) ^{2}
\end{equation}
with $V_{max}$ and $r_{max}$ being the maximum rotational velocity of the halo and the radius at which this velocity peak is located, respectively. This approach provides an alternative and faster route to compute concentrations compared to directly fitting individual radial density profiles with a Navarro-Frenk-White \citep[NFW, ][]{NFW} shape and has proven to be accurate in all circumstances where the relation between density and velocity perturbations does not deviate from its standard Newtonian form, while this correspondence does not hold anymore for e.g. modified theories of gravity (see again BV18 for a direct example in the case of $f(R)$ gravity). With such concentrations catalogs at hand, we %have
compute a binned concentration-mass relation by adopting the same binning already employed for the halo mass function displayed in Fig.~\ref{fig:HMF} and averaging over all halos belonging to each mass bin. The results are shown in Fig.~\ref{fig:conc}, where we display in the left panel the absolute concentrations while ratios to the standard Gaussian case are displayed in the right panel.

As one can notice from the figures, all models tend to recover the standard Gaussian prediction at the largest masses available in our sample, while deviations appear at smaller masses. In particular, the models with the largest $|f_{\rm NL, max}^{\rm loc}|=10^{4}$ have relative deviations up to $\approx 50\%$ compared to the Gaussian realization at the lowest end of the probed mass range. We overplot again for a direct comparison the suppression of halo concentrations for a WDM model with $m_{\rm WDM}=0.4$ keV using once again data from the WDM simulations of BV18. Interestingly, also in this case the WDM model  shows a deviation from $\Lambda $CDM with a similar shape to the case of our non-Gaussian model with $f_{\rm NL, max}^{\rm loc}=-10^{4}$, but with a significantly different amplitude. Basically, what we find is that WDM suppresses much more significantly the abundance of low-mass halos compared to our non-Gaussian model (as previously shown in Fig.~\ref{fig:HMF}), while at the same time has a much weaker effect on halo concentrations.

\subsection{Subhalo Mass Function}

\begin{figure}
\center
\includegraphics[width=0.7\columnwidth]{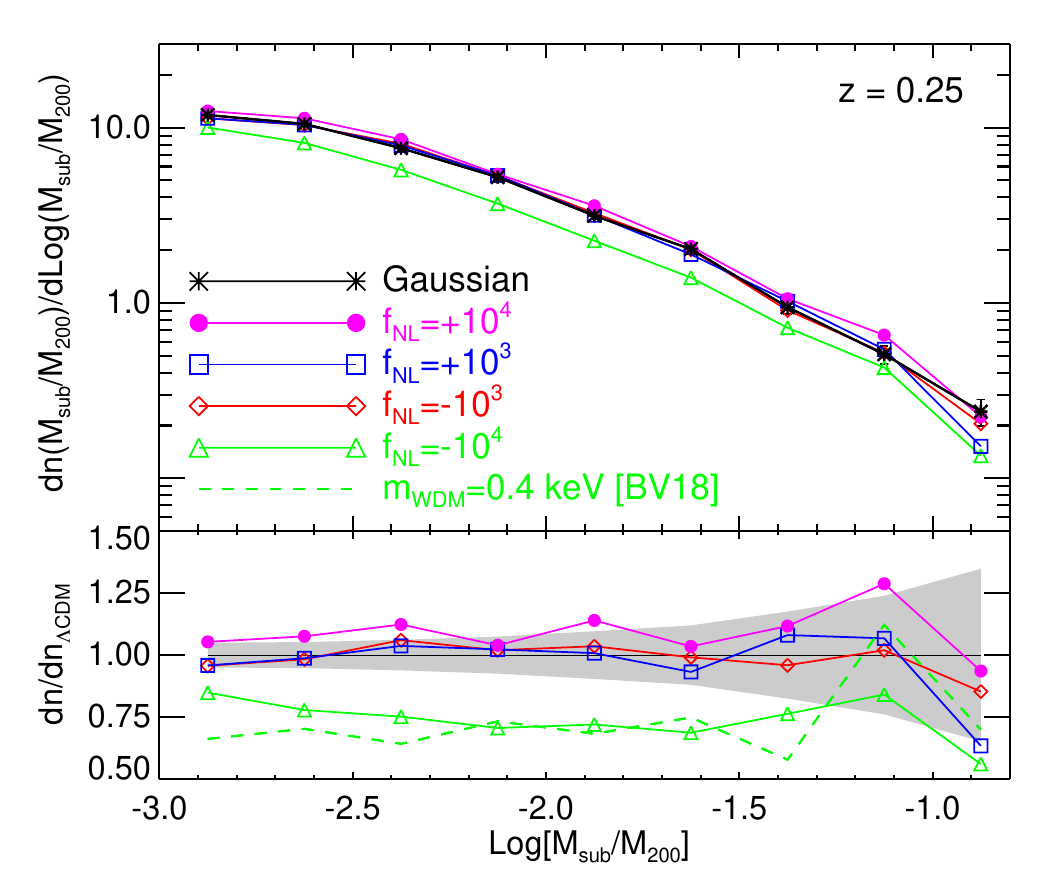}
\caption{{ The subhalo mass function of the different models ({\em top}) and its ratio to the reference cosmology ({\em bottom}) at $z=0$. The (barely visible) error bars in the {\em top} panel around the Gaussian curve represent the Poissonian error on the measured abundance while the grey-shaded area in the {\em bottom} panel represents the propagation of such Poissonian error on the relative deviation from the Gaussian case. Again, in the {\em bottom} panel we overplot the suppression of the subhalo mass function obtained for a Warm Dark Matter particle candidate with $m_{\rm WDM}=0.4$ keV by \cite{Baldi_Villaescusa-Navarro_2018}.}}
\label{fig:subHMF}
\end{figure}

Another relevant statistics for the models under investigation is  
the abundance of  substructures orbiting around a main dark matter halo. This is encoded in the subhalo mass function, defined as the number of subhalos of mass $M_{\rm sub}$ that are gravitationally bound to a main halo of virial mass $M_{200}$, as a function of the mass ratio $M_{\rm sub}/M_{200}$. We 
compute the subhalo mass function for all our cosmological models and compare it to the standard Gaussian case. The results are shown in Fig.~\ref{fig:subHMF} where we plot in the upper panel the absolute subhalo mass function and in the lower panel its ratio to the standard (Gaussian) case. The error bars (barely visible) around the Gaussian curve in the top panel indicate the Poissonian error on the subhalo counts in each mass ratio bin, while the grey-shaded region in the bottom panel shows the propagation of this error on the ratio, thereby providing an estimate for the statistical significance of deviations from the Gaussian reference model.

As the plot shows, the models with the lower value of $|f_{\rm NL, max}^{\rm loc}|=10^{3}$ are still consistent with the Gaussian case within statistical errors. For the $f_{\rm NL, max}^{\rm loc}=+10^{4}$ the comparison indicates a mild trend of enhancement of the subhalo abundance, even though for several bins of the mass ratio the standard Gaussian expectation is recovered. 
On the other hand, the $f_{\rm NL, max}^{\rm loc} = -10^{4}$ shows a very clear suppression of the abundance of substructures, with about $25-30$\%  fewer subhalos than the Gaussian realization for all mass ratios $\log [M_{\rm sub}/M_{200}] < -1.6$. Once again, this feature is shared by WDM scenarios \citep[actually being one of the reasons behind the widespread interest in WDM models as a possible solution to the so-called {\em missing satellite problem}, see e.g.][]{Moore_etal_1999,Klypin_etal_1999}, and also in this case we take the opportunity to compare it directly with the results of BV18 by overplotting the subhalo mass function suppression for $m_{\rm WDM}=0.4$ keV as a dashed green curve in the bottom panel of Fig.~\ref{fig:subHMF}. As the comparison shows, the reduction of halo substructures is similar in the two cases, showing that the $f_{\rm NL, max}^{\rm loc}=-10^{4}$ could address with similar effectiveness the so-called {\em satellite problem} of CDM, while affecting much less the abundance of main halos.
{We notice that the effect of positive and negative non-Gaussianity on the subhalo mass function is not symmetric, with the negative non-Gaussianity having a stronger impact for fixed $|f_{\rm NL,max}^{\rm loc}|$. As the subhalo mass function is sensitive to highly non-linear processes -- such as the tidal disruption of orbiting satellites that may in turn be influenced by a change in the halo concentrations -- we would  not expect to observe a symmetric deviation in this case.}

\subsection{Stacked halo density profiles}
\label{sec:halo_profiles}

\begin{figure*}
\includegraphics[width=\textwidth]{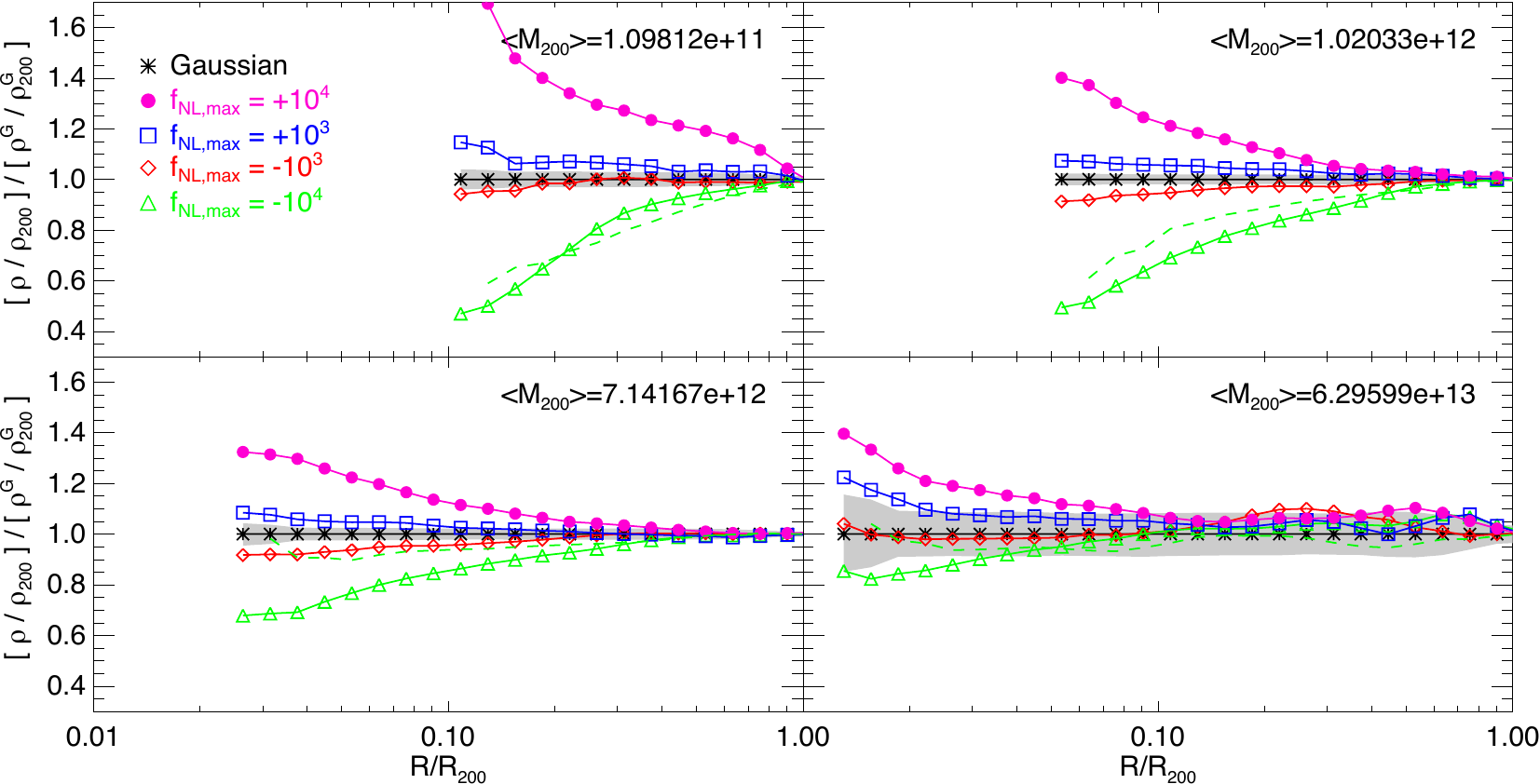}
\caption{{The ratio of the stacked halo density profiles to the Gaussian case. {Different panels refer to different bins in halo mass: the corresponding mean mass value is reported in each panel.} for various halo masses. The grey-shaded region represents the $1-\sigma$ statistical confidence region around the reference model, computed through a bootstrap resampling technique. We overplot again as a dashed green curve the suppression of the density profile for a Warm Dark Matter particle candidate with $m_{\rm WDM}=0.4 keV$ as obtained by \cite{Baldi_Villaescusa-Navarro_2018}.}}
\label{fig:profiles}
\end{figure*}

We 
compute the stacked spherically-averaged halo density profiles  from 100 randomly sampled halos within each of the 10 mass bins adopted for the halo mass function analysis discussed above. For the highest-mass bin the total number of halos is lower than 100 and we use all available halos.

Individual profiles are computed by binning particles in 30 logarithmically-equispaced radial shells centered on the most bound particle of the halo after rescaling the individual radial coordinates in units of the halo virial radius $R_{200}$. Stacked profiles are then expressed as a function of the radial distance from the center in units of the virial radius, and normalized in amplitude to the same value at $R/R_{200}=1$. 

This allows us to directly compare the shape of the profiles by simply plotting the ratio of the stacked profiles to the reference Gaussian realization. By construction, all ratios will converge to unity at $R/R_{200}=1$. We show the results of the comparison in Fig.~\ref{fig:profiles}, for 
four selected mass bins, where we have highlighted as a grey shaded area the 1-$\sigma$ statistical confidence region around the reference model, computed through a {\em bootstrap} resampling technique with 1000 re-samples of the 100 individual profiles. 

As the figure shows, the models with the lower value of $|f_{\rm NL, max}^{\rm loc}|=10^{3}$ are generally marginally consistent with the standard Gaussian reference model, with deviations never exceeding $\approx 10\%$ except for the very central regions (about a few percent of the virial radius) of the most massive halos, where the $f_{\rm NL,max}^{\rm loc}=+10^{3}$ scenario shows a steep rise of the density profile. Besides occurring very close to the resolution limit of our simulations, thus being prone to numerical artifacts, this feature does not significantly alter the overall mass distribution of such massive halos, as confirmed by the very mild impact of the same cosmological model on the halo concentrations at comparable masses (see Fig.~\ref{fig:conc} above).

On the other hand, the models with the largest value of $|f_{\rm NL, max}^{\rm loc}|=10^{4}$ show larger deviations from the standard Gaussian profiles at all masses, and over a relevant fraction of the virial radius of the halos, with the positive $f_{\rm NL}$ model resulting in a significant steepening of the profiles and the negative $f_{\rm NL}$ one producing a significant flattening instead. The maximum deviation occurs for small-mass halos, with the former reaching a $\gtrsim 70\%$ enhancement and the latter a $\gtrsim 50\%$ suppression at $1/10$th of the virial radius, while progressively larger halos show milder deviations at the same radial position.

It is particularly interesting to focus again on the $f_{\rm NL, max}^{\rm loc}=-10^{4}$ model, which shows a suppression of the density profiles (especially for the low-mass bins)  that closely resembles the expected effect of a WDM particle candidate. 

To quantitatively assess this similarity, as done for other observables, we overplot in the four panels of Fig.~\ref{fig:profiles}  the density profiles relative to   $\Lambda$CDM obtained for $m_{\rm WDM}=0.4$ keV by BV18, where individual halo density profiles extracted from BV18 simulation snapshots have been binned with the same procedure detailed above. As one can see from the plot, the density suppression of the $f_{\rm NL, max}^{\rm loc}=-10^{4}$ model closely resembles that of WDM at the smallest masses, but while for larger masses the WDM effect rapidly vanishes, the non-Gaussian model suppression remains significant also for galaxy- and group-sized halos. This shows once again that, for a comparable impact on the density power spectrum, our scale-dependent hyperbolic non-Gaussian models have a more pronounced effect on the structural properties of nonlinear collapsed halos with respect to WDM.

\subsection{Stacked void density profiles}
\label{sec:void_profiles}

\begin{figure*}
\includegraphics[width=\textwidth]{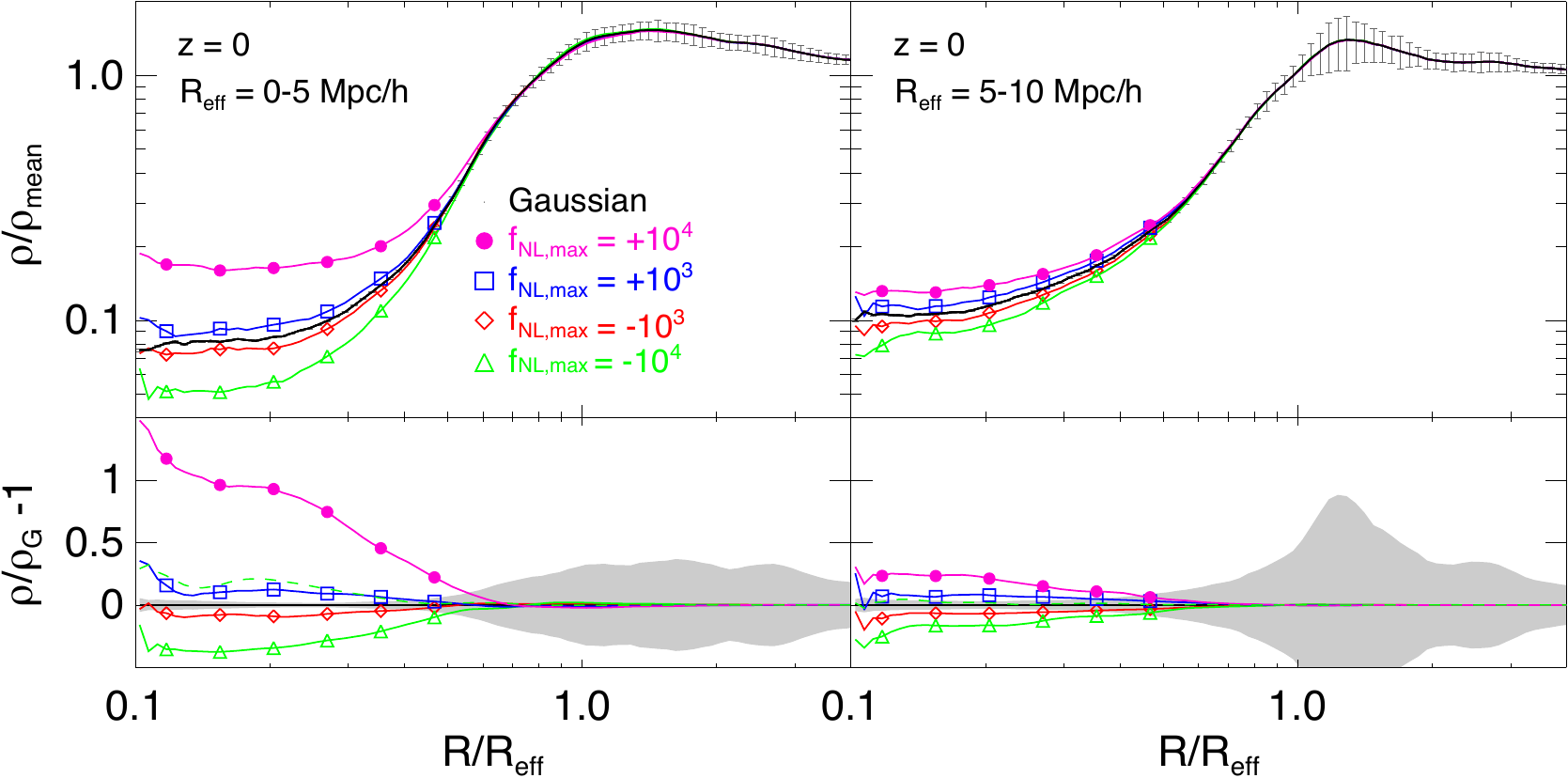}
\caption{{The stacked void density profiles (in the {\em top} panels) for two bins of void effective radius ($0-5$ Mpc$/h$ on the {\em left} and $5-10$ Mpc$/h$ on the {\em right}) and the corresponding relative difference (in the {\em bottom} panels) with respect to the Gaussian case for all the models under investigation. The dashed green curve in the {\em bottom} panel shows the behavior of a WDM particle candidate with $m_{\rm WDM}=0.4$ keV as obtained from the simulations of \cite{Baldi_Villaescusa-Navarro_2018}.  {The
grey-shaded regions represents the $2\sigma$ confidence region based on a bootstrap computation of the standard deviation of the average
profiles.}}}
\label{fig:void_profiles}
\end{figure*}	

As a last observable for our analysis, we consider the mass density distribution around cosmic voids, identified through the void finding procedure described above in Section~\ref{voidfind}. Also in this case, it is well known \citep[see e.g.][and BV18]{Yang_etal_2015} that a WDM particle candidate would give rise to a distinctive deviation from the predictions of standard CDM, namely a shallower density profile of cosmic voids with a higher central density than their CDM counterparts. This effect is more pronounced for small voids, as it is to be expected from the small-scale cutoff of the WDM primordial density perturbations spectrum. It is therefore natural to  
investigate 
a possible further degeneracy with the scale-dependent non-Gaussianity models
considered in the present work. 

To this end, we 
compute
the average void density profiles for voids with effective radius $R_{\rm eff}$ in the range $0-5$ Mpc$/h$ and $5-10$ Mpc$/h$, by stacking the density profiles of 100 randomly selected voids for each of these two bins of $R_{\rm eff}$. The resulting stacked profiles are shown in the {\em left} and {\em right} plots of Fig.~\ref{fig:void_profiles}, respectively. In the two plots, the top panels display the density profiles, with the (barely visible) error bars on the standard $\Lambda $CDM curves representing the statistical (Poissonian) errors on the mean based on the number of member voids, while the bottom panels show the relative difference with respect to the reference model, with the shaded areas indicating the 2-$\sigma$ confidence region computed through a {\em bootstrap} resampling technique with 1000 re-samples of the 100 individual profiles, as already done for the halo density profiles discussed in Section~\ref{sec:halo_profiles} above. 

As one can see from the plots, the effect of primordial non-Gaussianity on the void profiles is clearly visible for both void sizes, being more pronounced for the smaller ones. More specifically, while the weaker non-Gaussianity models with $|f_{\rm NL, max}^{\rm loc}|=10^{3}$ show deviations of the inner void density of the order of $10-20\%$ in both bins of effective radius, the $|f_{\rm NL, max}^{\rm loc}|=10^{4}$ models result in deviations reaching $30\%$ for the larger voids and exceeding $100\%$ for the smaller ones. 

Interestingly, we find that positive non-Gaussianity translates into shallower density profiles and an increase of the central void density, similarly to the case of WDM discussed by 
\cite{Yang_etal_2015} and BV18, while negative non-Gaussianity gives rise to the opposite trend, i.e. a steeper density profile and a lower central void density, which is a common feature of e.g. Modified Gravity \citep[see e.g.][ and BV18]{Cai_Padilla_Li_2015} and interacting Dark Energy models \citep[][]{Pollina_etal_2016}. 

This trend  as a function of $f_{\rm NL}$ is opposite to that we observed for the properties of overdense collapsed structures:
for  overdense  regions,  negative $f_{\rm NL}^{\rm loc}$  has a qualitatively (and in most cases also quantitatively) effect similar to that of WDM. For underdense regions instead the WDM trend is mimicked by the positive $f_{\rm NL}^{\rm loc}$ models.

More quantitatively, the effect on the void profile of a WDM with particle mass of $m_{\rm WDM}=0.4$ keV matches that of the weak non-Gaussian scenario with $f_{\rm NL, max}^{\rm loc}=+10^{3}$ (see dashed green line in Fig.~\ref{fig:void_profiles} extracted from the simulations of BV18). For the other probes discussed above the match has always been with the strongest negative non-Gaussian realization $f_{\rm NL, max}^{\rm loc}=-10^{4}$ for this value of $m_{\rm WDM}$.

Therefore, the comparison or joint analsyis of statistics related to the structural properties of collapsed halos  and  the density profiles of small cosmic voids  provides a further way to disentangle WDM from our scale-dependent non-Gaussianity models.

It is also interesting to notice that the $f_{\rm NL, max}^{\rm loc}=-10^{4}$ scenario is simultaneously capable of mimicking WDM in those aspects that may  alleviate  
the so-called {\em small-scale crisis of CDM} (reduced abundance of halo satellites, see Fig.~\ref{fig:subHMF}, suppression of the halo concentration, see Fig.~\ref{fig:conc}, and shallower density profiles of small halos, see Fig.~\ref{fig:profiles}) and to empty
cosmic voids more efficiently than the standard Gaussian reference model, thereby possibly addressing the so-called {\em void problem} \citep[see][]{Peebles_2001}.

\section{Conclusions}
\label{sec:conclusions}

We have presented the results of a suite of cosmological N-body simulations in the standard $\Lambda $CDM model with initial conditions featuring a scale-dependent primordial non-Gaussianity of the {\em local} type. More specifically, we consider a shape of the $f_{\rm NL}^{\rm loc}(k)$ function characterized by a power-law dependence on scale at large scales, normalized to be consistent with current CMB constraints \citep[][]{Planck:2019kim} at the corresponding pivot scale $k_{0}=0.05$ Mpc$^{-1}$, followed by a hyperbolic tangent term that determines a saturation of the $f_{\rm NL}^{\rm loc}(k)$ function to a maximum value $f_{\rm NL, max}^{\rm loc}$ that can be set as a free parameter of the model. This setup is devised in order to allow for very large values of the non-Gaussianity parameter at small scales where non-linear structure formation shapes the properties of a variety of collapsed structures (ranging from galaxies to cosmic voids) while ensuring 
consistency with current observational bounds at larger scales. The saturation at a value $f_{\rm NL,max}^{\rm loc}$ 
avoids the small-scale divergence of the $f_{\rm NL}^{\rm loc}$ in a 
pure power-law scale-dependence.

By means of the \texttt{PNGRun} initial conditions  code \citep[][]{Wagner_Verde_Jimenez_2012} we have generated initial conditions for five total models including a reference Gaussian scenario and two pairs of scale-dependent non-Gaussianity models characterized by $f_{\rm NL, max}^{\rm loc}=\pm 10^{3}$ and $f_{\rm NL, max}^{\rm loc}=\pm 10^{4}$ in a periodic cosmological box of $100$ Mpc $h^{-1}$ per side filled with $512^3$ particles.  Such initial conditions  have then  been evolved down to $z=0$ with the \texttt{Gadget-3} simulations code, and the resulting matter distribution  analyzed in terms of a wide range of observables. 

Our main findings can be summarised as follows:
\begin{itemize}
\item[$\star $]  The shape of the large-scale matter distribution appears very similar in all the simulations, showing that even the most extreme models with $f_{\rm NL, max}^{\rm loc}=\pm 10^{4}$ do not significantly alter the topology of the cosmic web at large scales;
\item[$\star $] The nonlinear matter power spectrum extracted from the simulations snapshots shows scale-dependent deviations from the reference Gaussian scenario in the direction of an enhanced (suppressed) amplitude of the non-linear power at small scales beyond $k\gtrsim 2\, h/$Mpc for models with positive (negative) non-Gaussianity. Such deviations reach an amplitude of about $8-10\%$ at the smallest scales probed by our simulations ($k\approx 20\, h$/Mpc) for the milder models with $f_{\rm NL, max}^{\rm loc}=\pm 10^{3}$, while for the more extreme scenarios with  $f_{\rm NL, max}^{\rm loc}=\pm 10^{4}$ the deviation reaches values of about $15-20\%$ at the same scales. In the recent paper by \cite{Stahl:2024stz}, investigating similar scenarios to the one considered in this work, such suppression has been claimed to provide a possible solution to the $S_{8}$ tension. Interestingly, we find that the shape and the amplitude of such deviations for the models with negative non-Gaussianity closely match those predicted for Warm Dark Matter particle candidates using the fitting function of \cite{Viel_etal_2013} calibrated on high-resolution simulations. More specifically, the model with $f_{\rm NL, max}^{\rm loc}=- 10^{3}$ shows a deviation consistent with a Warm Dark Matter particle mass of $m_{\rm WDM}\approx 0.6$ keV, while the model with $f_{\rm NL, max}^{\rm loc}= - 10^{4}$ is found to reproduce the expected suppression for $m_{\rm WDM}\approx 0.4$ keV. Remarkably, this intriguing observational degeneracy between our primordial non-Gaussianity models and Warm Dark Matter phenomenology appears also in other observables, as outlined below.
\item[$\star $] The abundance of halos encoded by the Halo Mass Function shows very mild and symmetric deviations with respect to the Gaussian case for the two models with $f_{\rm NL, max}^{\rm loc}=\pm 10^{3}$, with the positive (negative) non-Gaussianity model showing a few percent suppression (enhancement) of the halo abundance at masses below $\approx 1.7\times 10^{12}$ M$_{\odot}/h$ followed by an enhancement (suppression) of the same magnitude for larger masses. The situation is quite different for the models with $f_{\rm NL, max}^{\rm loc}=\pm 10^{4}$, for which the symmetry is lost and the positive (negative) non-Gaussianity determines an enhancement (suppression) over the whole range of masses probed by our halo sample, with the negative model recovering the abundance of the Gaussian case only at the highest mass bin around $7\times 10^{13}$ M$_{\odot}/h$. In this case, the observed behavior is both qualitatively and quantitatively different from the case of the WDM particle mass that was found to match the non-linear power spectrum, with the non-Gaussian models showing a stronger suppression at large masses but a much milder suppression at small masses compared to WDM.
\item[$\star $]  The halo bias shows a mild enhancement (suppression) with respect to the Gaussian case for the positive (negative) non-Gaussianity models only for the most extreme scenarios with $f_{\rm NL, max}^{\rm loc}=\pm 10^{4}$, with an amplitude of $\approx 5-10\%$, while the milder models show basically no deviation from the Gaussian case.
\item[$\star $] The concentration-mass relation computed from our halo sample shows an almost mass-independent enhancement (suppression) for the positive (negative) non-Gaussian models with $f_{\rm NL, max}^{\rm loc}=\pm 10^{3}$, with a relative deviation from the Gaussian case of $\approx 10\%$. On the other hand the more extreme models with $f_{\rm NL, max}^{\rm loc}=\pm 10^{4}$ show mass-dependent deviations going in the same directions, with small-mass halos being more strongly affected with deviations reaching $\approx 50\%$ at the low-mass end of our sample and decreasing down to $\approx 10\%$ at the high-mass end. By comparing these results with the deviation obtained from WDM simulations with the same particle mass showing degeneracy in the non-linear power spectrum ($m_{\rm WDM}=0.4$ keV) we notice that the shape of the deviation as a function of mass is very similar for the two scenarios, but the amplitude appears to be about twice as large for the non-Gaussian model with respect to the WDM one. Interestingly, this result shows that these particular primordial non-Gaussianity models can suppress halo concentrations more efficiently with respect to even extreme (and already ruled out) Warm Dark Matter scenarios.
\item[$\star $] We also tested the impact of these different models on the SubHalo mass function computed from our halo and subhalo catalogs. We observe that the models with $f_{\rm NL, max}^{\rm loc}=\pm 10^{3}$ appear to be consistent with the reference Gaussian scenario within statistical errors, while the models with $f_{\rm NL, max}^{\rm loc}=\pm 10^{4}$ show some significant deviations from the Gaussian case. In particular, the negative non-Gaussianity model shows a $\approx 25\%$ suppression of the abundance of subhalos which appears again comparable with the one obtained from WDM simulations with the same particle mass showing degeneracy in the non-linear power spectrum ($m_{\rm WDM}=0.4$ keV). Therefore, we showed here for the first time how these scale-dependent primordial non-Gaussianity models may be as effective as very extreme WDM models in suppressing the number of satellites in gravitationally bound structures.
\item[$\star $] We tested the impact of our primordial non-Gaussianity models on the density profile of halos by dividing our halo sample into four mass bins and computing the stacked density profiles in units of the virial radius $R_{200}$ for 100 randomly selected halos in each mass bin. We showed that the positive (negative) non-Gaussianity models determine a steeper (shallower) density profile of halos for all the four mass bins, with a stronger effect appearing for smaller halos and for larger values of $|f_{\rm NL, max}^{\rm loc}|$. Also in this case, we compared the result observed for the $f_{\rm NL, max}^{\rm loc}=- 10^{4}$ model with the suppression obtained from a WDM simulation with the same particle mass showing degeneracy in the non-linear power spectrum ($m_{\rm WDM}=0.4$ keV), showing that -- quite remarkably -- the degeneracy between the two models holds also for the halo density profiles of small mass halos, corresponding to our first mass bin with an average virial mass of $<M_{200}>\approx  10^{11}$ M$_{\odot }/h$, while for the second mass bin with $<M_{200}> \approx 10^{12}$ the suppression given by WDM is slightly weaker compared to the non-Gaussian scenario, still showing the same shape. This trend is confirmed also for the higher mass bins, with the WDM model recovering full consistency with standard Gaussian $\Lambda $CDM profiles at the largest mass bin, while the non-Gaussian model shows a residual suppression over the whole mass range. This is again a quite remarkable result, as it shows that our class of non-Gaussian models can simultaneously impact satellite abundances and halo density profiles in a similar way as a very extreme WDM model, thereby possibly providing a completely new mechanism to address the longstanding small-scale problems of CDM without resorting to baryonic effects.
\item[$\star $] As a final observational statistics of our models, we have computed the stacked void density profiles starting from the corresponding void catalogs. We showed that positive (negative) non-Gaussianity results in shallower (steeper) void density profiles, with small voids ($R_{\rm eff }\leq 5$ Mpc$/h$) being more strongly affected than larger voids ($R_{\rm eff}=5-10$ Mpc$/h$). Interestingly, this is the opposite behavior of what is observed for Warm Dark Matter \citep[see e.g.][]{Yang_etal_2015,Baldi_Villaescusa-Navarro_2018}, as we also confirm, showing that the degeneracy we observed in several other quantities between negative scale-dependent non-Gaussianity in the form of Eq.~\ref{eq:hyperbolicfNL} and a WDM particle candidate is broken by cosmic voids properties.
\end{itemize}

To conclude, we highlighted for the first time an intriguing similarity between the  effects of {running non-Gaussianity models with $f_{\rm NL}^{\rm loc
}(k)<0$} on some of the tested observables and those determined {on the same observables} by Warm Dark Matter particle candidates, which holds both at the qualitative and quantitative level as a function of scale and mass. Although this degeneracy does not appear in all the statistics that we investigated, and can be possibly broken by considering statistics of both high-density regions/halos and voids, we found that a scale-dependent non-Gaussianity model of the kind presented in this work may simultaneously suppress the abundance of satellites and the central overdensity of halo profiles, thereby possibly providing a new avenue to address some of the longstanding small-scale issues of the CDM paradigm.

\section*{Acknowledgments}
MB is supported by the project ``Combining Cosmic Microwave Background and Large Scale Structure data: an Integrated Approach for Addressing Fundamental Questions in Cosmology", funded by
the MIUR Progetti di Ricerca di Rilevante Interesse
Nazionale (PRIN) Bando 2017 - grant 2017YJYZAH.
MB acknowledges the use of computational resources provided by the INFN-InDark project at CINECA and by the Open Physics Hub at the Bologna University. LV and EF acknowledge Center of Excellence Maria de Maeztu 2020-2023? award to the ICCUB (CEX2019-000918-M funded by MCIN/AEI/10.13039/501100011033) and project PID2022-141125NB-I00 MCIN/AEI. EF acknowledges the support from ``la Caixa" INPhINIT Doctoral Fellowship (ID 100010434, code LCF/BQ/DI21/11860061). The work of Francisco Villaescusa-Navarro is supported by the Simons Foundation. DK is supported by the Science and Technology Facilities Council (UK) grant number 
ST/X000931/1. GJ acknowledges support from the ANR LOCALIZATION project,
grant ANR-21-CE31-0019 / 490702358 of the French Agence Nationale de la Recherche. 
AR acknowledges support from PRIN-MIUR~2020 METE, under contract no. 2020KB33TP. ML acknowledges support by the MIUR Progetti di Ricerca di Rilevante Interesse Nazionale (PRIN) Bando 2022 - grant 20228RMX4A. LM acknowledges the financial contribution from the grant PRIN-MUR 2022 20227RNLY3 ``The concordance cosmological model: stress-tests with galaxy clusters" supported by Next Generation EU and from the grants ASI n.2018-23-HH.0 and n. 2024-10-HH.0 ``Attivit\`a scientifiche per la missione Euclid -- fase E"

%*****************************************************************************
\bibliographystyle{JHEP}
\bibliography{baldi_bibliography}

\end{document}